\documentclass[prb,superscriptaddress,longbibliography,twocolumn]{revtex4-1}
\usepackage{geometry}
\usepackage{amsmath}
\usepackage{amssymb}
\usepackage{graphicx}
\usepackage{hyperref}
\usepackage{braket}
\usepackage{bm}

\usepackage{tikz}
\usepackage{environ}

\makeatletter
\newsavebox{\measure@tikzpicture}
\NewEnviron{scaletikzpicturetowidth}[1]{%
  \def\tikz@width{#1}%
  \begin{lrbox}{\measure@tikzpicture}%
  \BODY
  \end{lrbox}%
  \pgfmathparse{#1/\wd\measure@tikzpicture}%
  \BODY
}
\makeatother

\def\G{\Gamma}
\def\a{\alpha}
\def\b{\beta}
\def\bk{{\bf k}}

\def\bq{{\bf q}}
\def\br{{\bf r}}

\begin{document}

\title{Signatures of Quantum Spin Liquid in Kitaev-like Frustrated
  Magnets}

\author{Matthias Gohlke}
\affiliation{Max-Planck-Institut f\"ur Physik komplexer Systeme, 
  01187 Dresden, Germany}

\author{Gideon Wachtel} 
\affiliation{Department of Physics, University of Toronto, Toronto,
  Ontario M5S 1A7, Canada}

\author{Youhei Yamaji}
\affiliation{Department of Applied Physics, The University of Tokyo,
  Hongo, Bunkyo-ku, Tokyo, 113-8656, Japan}
\affiliation{JST, PRESTO, Hongo, Bunkyo-ku,
  Tokyo, 113-8656, Japan}

\author{Frank Pollmann}
\affiliation{Technische Universit\"at M\"unchen, 85747 Garching, Germany}

\author{Yong Baek Kim} 
\affiliation{Department of Physics, University of Toronto, Toronto,
  Ontario M5S 1A7, Canada}
\affiliation{Canadian Institute for Advanced Research/Quantum
  Materials Program, Toronto, Ontario MSG 1Z8, Canada}
\affiliation{School of Physics, Korea Institute for Advanced Study,
  Seoul 130-722, Korea}

\begin{abstract} 
  Motivated by recent experiments on $\alpha$-RuCl$_3$, we investigate
  a possible quantum spin liquid ground state of the honeycomb-lattice
  spin model with bond-dependent interactions.  We consider the
  $K-\Gamma$ model, where $K$ and $\Gamma$ represent the Kitaev and
  symmetric-anisotropic interactions between spin-1/2 moments on the
  honeycomb lattice.  Using the infinite density matrix renormalization group
  (iDMRG), we provide compelling evidence for the existence of quantum
  spin liquid phases in an extended region of the phase diagram. In
  particular, we use transfer matrix spectra to show the evolution of
  two-particle excitations with well-defined two-dimensional
  dispersion, which is a strong signature of quantum spin
  liquid. These results are compared with predictions from Majorana
  mean-field theory and used to infer the quasiparticle excitation
  spectra.  Further, we compute the dynamical structure factor using
  finite size cluster computations and show that the results resemble
  the scattering continuum seen in neutron scattering experiments on
  $\alpha$-RuCl$_3$.  We discuss these results in light of recent and
  future experiments.
\end{abstract}

\maketitle

\section{Introduction}

One of the hallmarks of quantum spin liquid is the existence of
fractionalized excitations\cite{balents_spin_2010}.  While it is
generally difficult to detect the dispersion of single-particle
excitations in fractionalized systems, the information about
two-particle and other multi-particle excitations is contained in the
dynamical spin structure factor measured in inelastic neutron
scattering.  If the ground state is a quantum spin liquid, the lower
boundary of multi-particle continuum should have a well-defined
dispersion. Many candidate materials for quantum spin liquid, however,
do not allow such scattering experiment due to unavailability of large
single crystal. Possible experiments are, therefore, quite often
limited to thermodynamic measurements such as specific heat and
susceptibility, as well as thermal transport. Hence it is difficult to
identify smoking-gun evidence for quantum spin liquid.

In this context, recent inelastic neutron scattering 
experiments\cite{banerjee_proximate_2016, banerjee_neutron_2017,j_wen_2017}
on $\alpha$-RuCl$_3$ provide valuable information about multi-particle
continuum in a putative quantum spin liquid
material\cite{yjkim_2014}.
$\alpha$-RuCl$_3$ is one of the candidate
materials that support the Kitaev interaction\cite{kitaev_anyons_2006}
between $j_{\rm eff}=1/2$ pseudo-spin moments, which is a
bond-dependent frustrated Ising interaction and arises from the
combination of correlation effects and spin-orbit
coupling\cite{jackeli_mott_2009,ybkim_2014,
rau_spin-orbit_2016,winter_models_2017,trebst_2017}. 
When only the Kitaev interaction is present, such a model
can be exactly solved\cite{kitaev_anyons_2006} and the ground state is
a quantum spin liquid.  On the other hand, other interactions are
generally present and they may drive a transition to a magnetically
ordered state\cite{rau_generic_2014, winter_challenges_2016,
  kim_crystal_2016, winter_breakdown_2017}.

The compound $\alpha$-RuCl$_3$ orders magnetically at low
temperatures\cite{johnson_monoclinic_2015, sears_magnetic_2015,
  little_antiferromagnetic_2017}, possibly due to the existence of
other interactions mentioned above. In spite of this, the dynamical
structure factor shows a continuum of excitations at high energies
both below and above the ordering temperature, which may be related to
a nearby quantum spin liquid\cite{gohlke_dynamics_2017}.  Elaborate
{\it ab initio} computations indicate that the dominant exchange
interactions are the Kitaev $K$ and symmetric-anisotropic $\Gamma$
interactions with small third neighbor $J_3$ Heisenberg
interaction\cite{winter_challenges_2016}. The dominance of $K$ and
$\Gamma$ interactions is also pointed out in studies of multi-orbital
Hubbard model\cite{Jian_Xin_Li_2016}. Furthermore, it is interesting
to notice that the $K-\Gamma$ model is proposed for metal-organic
frameworks with Ru$^{3+}$ or Os$^{3+}$
ions\cite{yamada_designing_2017}.  A previous theoretical
work\cite{catuneanu_realizing_2017} of exact diagonalization (ED) on
finite size clusters suggest that the $K-\Gamma$ model may host
quantum spin liquid phases in an extended region of the phase diagram
and small $J_3$ would drive a transition to the zig-zag order that is
seen in the experiment.  Such a study is naturally subject to finite
size effect and it is still difficult to nail down the precise nature
of the ground state.  In fact, the ED study mentioned above introduce
a spatial anisotropy in the Kitaev interaction to avoid strong finite
size effect.

In this work, we investigate the $K-\Gamma$ model using then infinite
density matrix renormalization group\cite{white_density_1992,
  mcculloch_infinite_2008, kjall_phase_2013}(iDMRG), where the system
is placed on an infinite cylinder.  We first study the ground state
energy, entanglement entropy, magnetization, and static structure
factor for the $K-\Gamma$ model. Similar to the previous ED study, we
find quantum paramagnetic ground states for ferro-like Kitaev
interaction for an extended region in the phase diagram. It is also
shown that the entanglement entropy in this region remains relatively
high.  In comparison, the entanglement entropy of a magnetically
ordered state for anti-ferro-like Kitaev interaction is quite small.

In order to test the existence of coherent excitations, we compute the
eigenvalues of the transfer matrix (TM) in the matrix product
wavefunction.  Using the mapping\cite{zauner_transfer_2015} between
the complex eigenvalues and low-energy excitation spectra as a
function of momentum, we identify the lower edge of the multi-particle
excitation continuum as a function of
momentum\cite{he_signatures_2017}.  From this, we find that the lower
edge of the excitation spectra, which also corresponds to the longest
correlation length in the system, moves coherently in momentum space
in a well-defined fashion as a function of $\Gamma/K$. This alone
tells us that these states are non-trivial and likely correspond to
quantum spin liquid phases. Furthermore, the TM spectrum exhibits an
emergent symmetry in its momentum dependence. Assuming that the lowest
momentum-dependent eigenvalues correspond to the lower boundaries of
the single- and multi-particle continuum in a quantum spin liquid, we
compare the results of the transfer matrix computations and the
two-particle spectra obtained in the Majorana mean-field theory for
the $K-\Gamma$ model. It is found that the same symmetry which emerges
in the TM spectrum is also a property of the Majorana spectrum in the
mean-field theory.  This suggests that the mean-field picture can
capture some essential features of the two-particle excitation
spectrum seen in the transfer matrix eigenvalues.

As pointed out in earlier works\cite{baskaran_exact_2007,
  knolle_dynamics_2014, knolle_dynamics_2015}, the dynamical spin
structure factor in the Kitaev model involves not just two-particle
spectra, but also the flux degrees of freedom. In fact, this makes it
difficult to extract the information about the single-particle
dispersion from the dynamical spin structure factor as the flux
degrees of freedom may still play important roles even in the presence
of the additional $\Gamma$ interaction. In contrast, the transfer
matrix eigenvalues computed here are directly related to the
convolution of the single-particle spectra.

In order to make a more direct contact with scattering experiment, we
compute the dynamical structure factor on a 24-site cluster.  It is
found that the low energy dynamical structure factor exhibit a
star-like feature in momentum space, just like what is seen in recent
neutron scattering experiments\cite{banerjee_neutron_2017} on
$\alpha$-RuCl$_3$.  We also demonstrate that these results are
consistent with the equal-time spin structure factor computed in the
Majorana mean-field theory.  We have further confirmed that the
magnetic-field dependence of the magnetization in the 24-site cluster
is consistent with the experimental results\cite{yadav_kitaev_2016,
  baek_evidence_2017, hentrich_large_2017, zheng_gapless_2017,
  sears_phase_2017, wolter_field-induced_2017, wang_magnetic_2017,
  winter_breakdown_2017} on $\alpha$-RuCl$_3$.  Taken together, all of
the results above suggest that $\alpha$-RuCl$_3$ may be close to a
quantum spin liquid described by the $K-\Gamma$ model with ferro-like
Kitaev interaction.

The remainder of the paper is organized as follows. We present the model
and iDMRG results in section~\ref{sec:iDMRG}. In section \ref{sec:TM}
we present and discuss the transfer matrix (TM) spectrum. A Majorana
based MFT is presented in section \ref{sec:MFT}, while results of the
ED calculation of the dynamic structure factor appear in section
\ref{sec:dynsf}. Details of the iDMRG and ED calculations, as well as
additional results are given in the appendices.

\section{\lowercase{i}DMRG study of the $K-\G$ model}
\label{sec:iDMRG}

\begin{figure}[t]
  \centering
  \includegraphics[width=\linewidth]{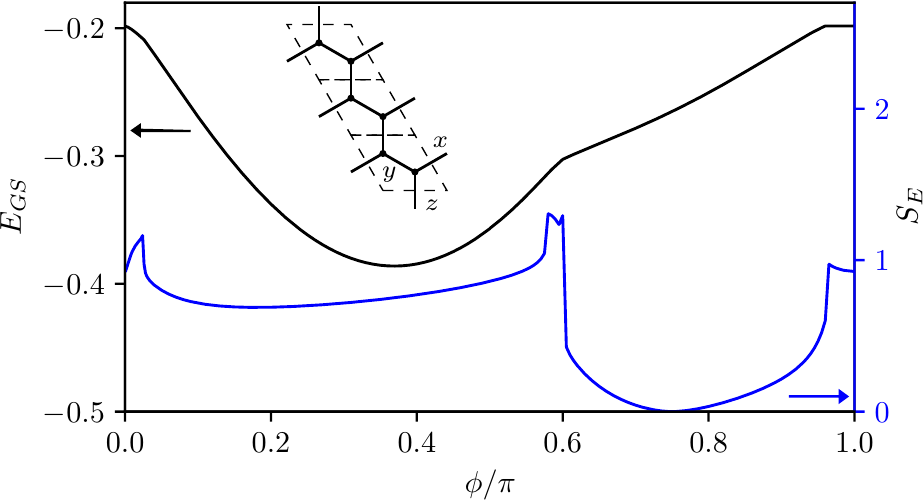}
  \caption{Ground state energy density, $E_{GS}$, of the $K-\G$ model,
    and the corresponding entanglement entropy, $S_E$, for a
    bipartition of the cylinder into two infinite cylinders, as
    determined using iDMRG for cylinders with circumference
    $L=6$. Kinks in $E_{GS}$ are indications of first order phase
    transitions. Inset shows the cylinder's three-unit-cell
    circumference.}
  \label{fig:EGSSE}
\end{figure}

\begin{figure}[t]
  \centering
  \includegraphics[width=\linewidth]{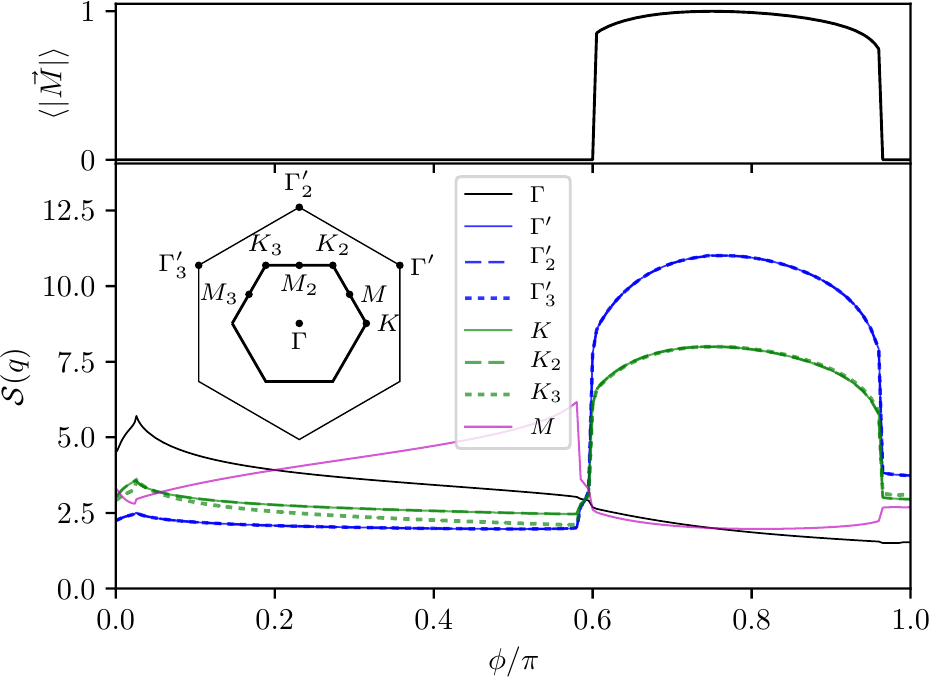}
  \caption{Staggered magnetization (top) and static spin structure
    factor (bottom), calculated using iDMRG. Inset: Brillouin zone
    with labeled positions of symmetry points.}
  \label{fig:MSSF}
\end{figure}

\begin{figure}[t]
  \centering
  \includegraphics[width=\linewidth]{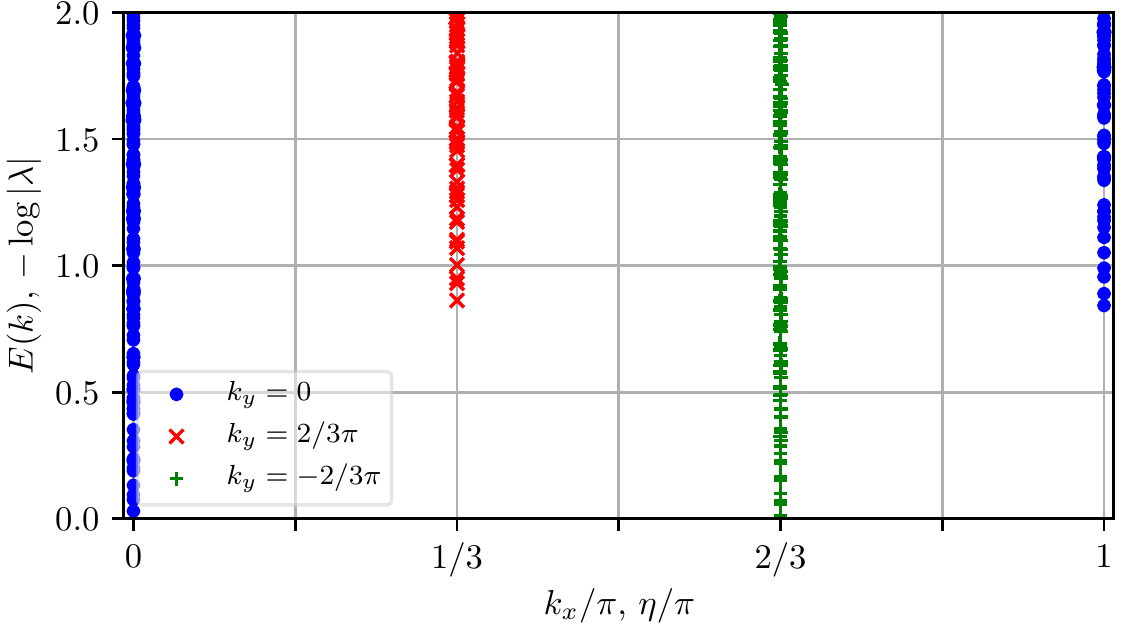}
  \caption{Transfer matrix spectrum $\lambda_i$ in the Kitaev limit ($\phi \rightarrow 0$) for $L=6$.  Plotted is
    the spectrum $E(k) = - \log |\lambda|$ with each point
    corresponding to a single eigenvalue $\lambda$, where $\lambda =
    |\lambda| e^{i\eta}$ and $\eta$ is identified with the momentum
    $k_x$ along the cylinder. $k_y$ denotes the transverse momentum
    obtained as a quantum number with respect to translation along the
    cylinder. See App. A for more details.
    }
  \label{fig:KTM}
\end{figure}
\begin{figure}[t]
  \centering
  \includegraphics[width=\linewidth]{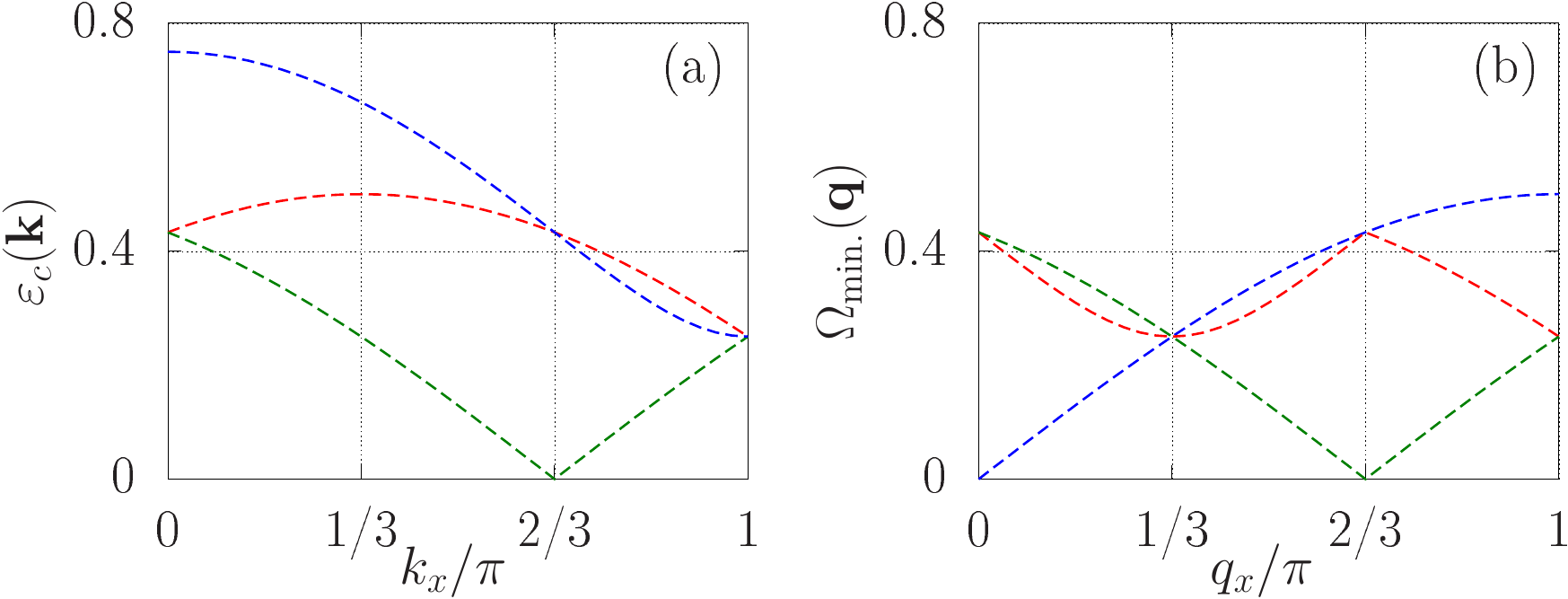}
  \caption{(a) Single Majorana fermion spectrum $\varepsilon_c(\bk)$
    for the isotropic Kitaev model, $K_x=K_y=K_z=-1$, on a
    cylinder with a three unit cell circumference. (b) Corresponding
    minimum energy for two-particle excitations, $\Omega_{\rm
      min.}(\bq)$.}
  \label{fig:Kminex2}
\end{figure}
\begin{figure*}[t!!]
  \centering
  \includegraphics[width=\linewidth]{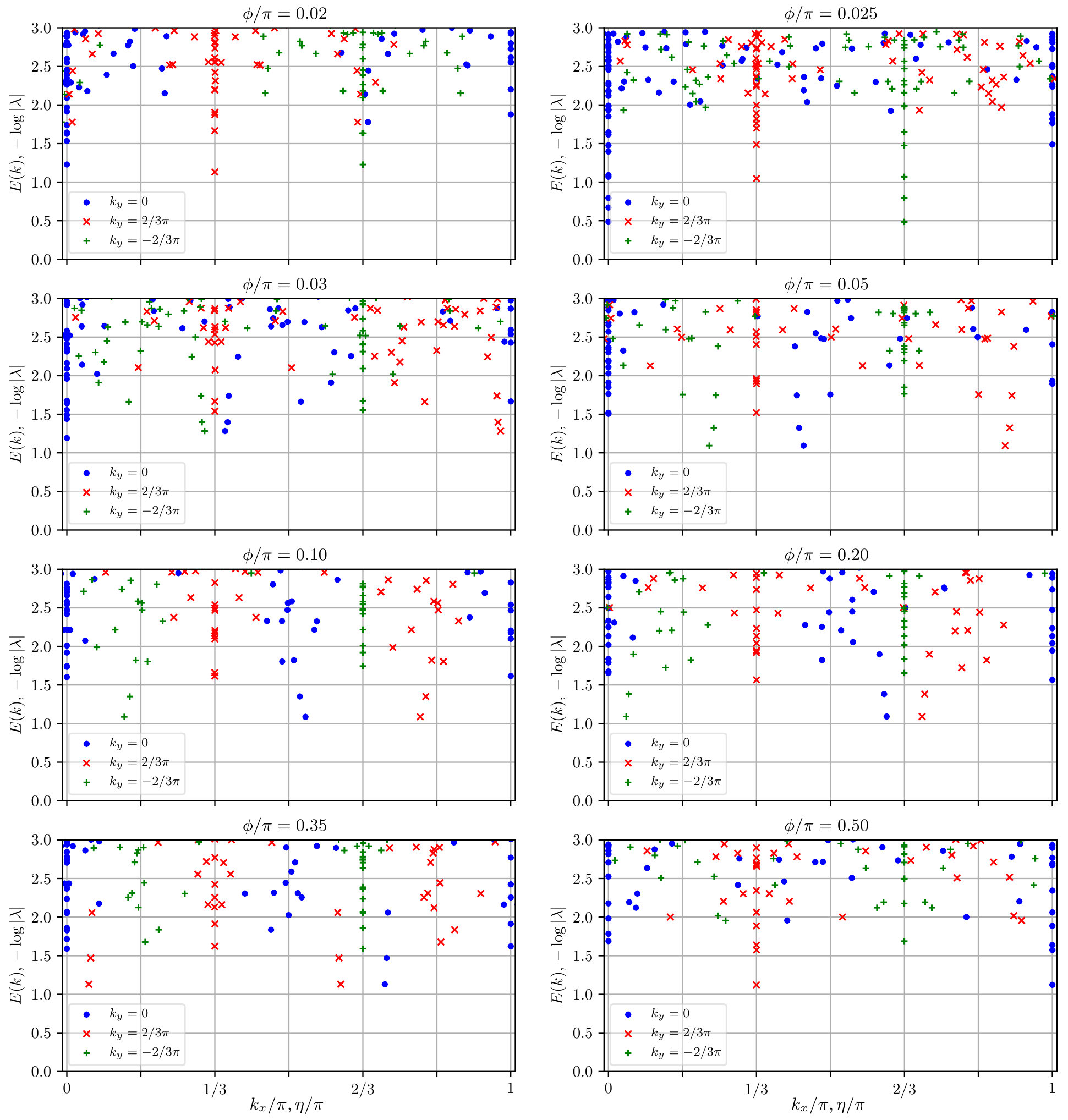}
  \caption{Same as Fig. \ref{fig:KTM} for different $\phi$.}
  \label{fig:TM}
\end{figure*}

The $K-\G$ Hamiltonian is given by 
\begin{eqnarray}
  \label{eq:Hb}
  H & = & \sum_{\braket{ij}\in x}K_xS_i^xS_j^x+\Gamma_x(S_i^yS_j^z+S_i^zS_j^y)
  \nonumber \\ & & + 
  \sum_{\braket{ij}\in y}K_yS_i^yS_j^y+\Gamma_y(S_i^xS_j^z+S_i^zS_j^x)
  \nonumber \\ & & + 
  \sum_{\braket{ij}\in z}K_zS_i^zS_j^z+\Gamma_z(S_i^xS_j^y+S_i^yS_j^x),
\end{eqnarray}
where $S_i^\a$ are spin-$1/2$ operators at site $i$ of a honeycomb
lattice. Unless otherwise noted, we consider here the isotropic case,
$K_\a=K,\G_\a=\G$. Throughout the following, $K$ and $\G$ are
parameterized using $\phi$ such that $K=-\cos\phi$ and
$\G=\sin\phi$. We use the iDMRG method with bond dimensions of up to
$\chi=400$ to obtain the ground state of this model on a narrow
infinite cylinder with a three unit cell circumference ($L=6$). In
Fig.  \ref{fig:EGSSE} we plot the ground state energy density,
$E_{GS}$, as a function of the parameter $0<\phi<\pi$. Starting from
the ferromagnetic Kitaev limit, $\phi=0$, $E_{GS}$ evolves smoothly
through the $\G$ limit, $\phi=\pi/2$. A discontinuity appears at
$\phi\approx 0.6\pi$ and again slightly before the anti-ferromagnetic
Kitaev limit, $\phi=\pi$. The two discontinuities are associated with
a transition into, and out of, a magnetically ordered vortex state,
which becomes an exact product state for $\phi=3\pi/4$. This is
evident in a plot of the entanglement entropy, $S_E$, also in Fig.
\ref{fig:EGSSE}, showing a vanishing $S_E$ at this point. Notice that
the entanglement entropy remains as high as that of the ferro-like
Kitaev limit in the entire region of $0 < \phi < 0.6 \pi$. Furthermore
iDMRG shows a finite staggered magnetization in $0.6\pi < \phi < 0.96\pi$, 
as well as an enhanced spin structure factor, Fig. \ref{fig:MSSF}, all
consistent with a magnetically ordered phase. The main question we
want to address here is: what is the nature of the ground state
outside of the magnetically ordered state?  The large entanglement and
lack of magnetic order suggest that the ground state in this region is
occupied by quantum spin liquid phases.  However, a small
discontinuity at $\phi\approx 0.025\pi$ in both the entanglement
entropy and the spin structure factor, raises the question whether
there exists a subtle transition between different kinds of spin
liquid phases. To address this issue, and to gain insight into the low
energy physics of the $K-\G$ model as $\phi$ is tuned from the Kitaev
to $\G$ limits, we turn to a detailed examination of the transfer
matrix spectrum, obtained from the ground state matrix product state (MPS).

\section{Transfer matrix spectrum}
\label{sec:TM}

\subsection{The Kitaev limit}

We begin by analyzing the transfer matrix spectrum, $E(k_x,k_y)$, of
the pure Kitaev model, shown in Fig. \ref{fig:KTM}, as a function of
the momentum along the cylinder, $k_x$. We were also able to resolve
the transverse momentum $k_y=0,\pm 2\pi/3$, which are depicted in the
figure by different colors (see Fig. \ref{fig:bz}). Here we use the
demonstrated correspondence\cite{zauner_transfer_2015} between the
complex eigenvalues of the transfer matrix and the excitation
spectrum, $E(k)$. Namely, given a TM eigenvalue
$\lambda_i=e^{-\epsilon_i+i\eta_i}$, the corresponding momentum (along
the infinite dimension) is given by $k_i\sim\eta_i={\rm
  arg}\lambda_i$, while the corresponding energy is given by
$E_i\sim\epsilon_i=-\ln|\lambda_i|$ (see appendix \ref{sec:appiDMRG}
for details). The Kitaev model is exactly solvable in terms of
Majorana fermions, and therefore it is possible to readily identify
the features in Fig. \ref{fig:KTM} with the known Majorana
excitations. The most prominent feature of the Majorana spectrum,
$\varepsilon_c(\bk)$, in the Kitaev model is the existence of two
gapless Dirac nodes at the corners of the Brillouin zone,
$K=(2\pi/3,-2\pi/3)$ and $K'=(-2\pi/3,2\pi/3)$ (see
Fig. \ref{fig:Kminex2}a). A continuum of excitations may thus be
obtained if multiple Majorana fermions are
excited. Fig. \ref{fig:Kminex2}b shows the minimum excitation energies
for the two particle excitation spectrum, as defined by
\begin{equation}
  \label{eq:minpolec}
  \Omega_{\rm min.}(\bq) = \min_{\bk}\left(|\varepsilon_c(\bq-\bk)|+
   |\varepsilon_c(\bk)|\right).
\end{equation}
Minima in $\Omega_{\rm min.}(q_x,q_y)$, as a function of $q_x$, in the
two particle spectrum, appear at $(0,0)$, $(\pi/3,2\pi/3)$ and
$(2\pi/3,-2\pi/3)$, which are consistent with the blue, red, and green
pillars shown in Fig. \ref{fig:KTM}a at these momenta.  We note,
however, that, at least in the pure Kitaev model, single-particle
excitations seem to appear in the TM spectrum as well (see Appendix
\ref{sec:appiDMRG} for details).

\subsection{The $K-\G$ model}
\label{sec:TM_KG}

\begin{figure}[t]
  \centering
  \includegraphics[width=\linewidth]{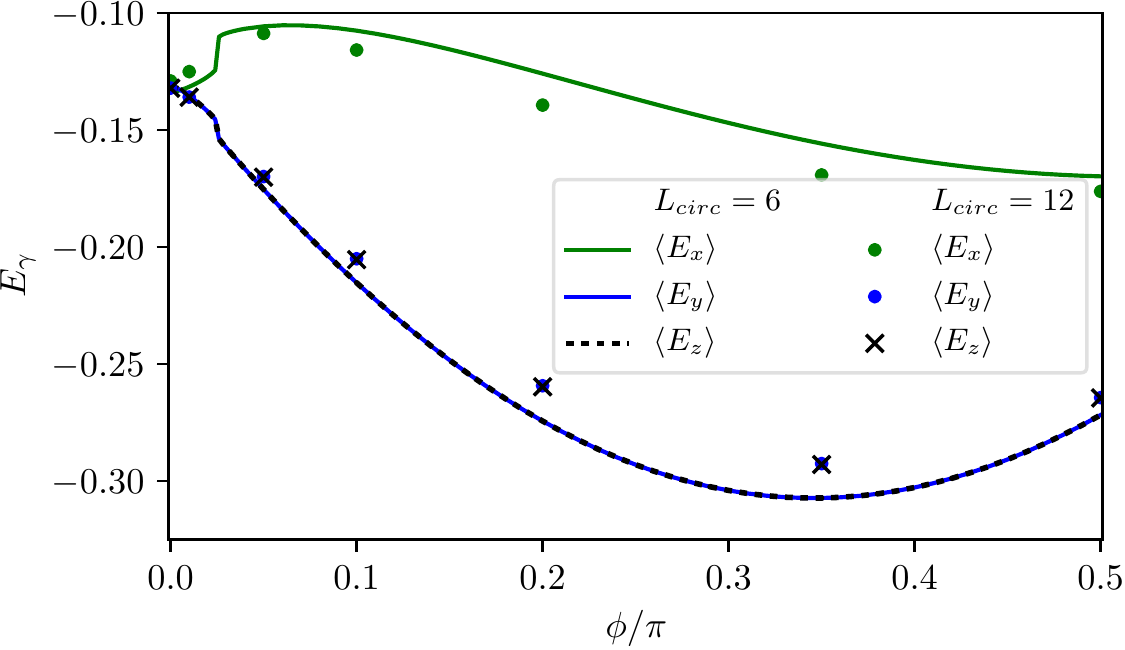}
  \caption{Energy density per bond, as obtained using iDMRG, for
    systems with a three ($L=6$) and six ($L=12$) unit cell
    circumference.}
  \label{fig:E_per_bond}
\end{figure}

Moving away from the exactly solvable Kitaev limit, we now turn to
analyze the TM spectrum of the $K-\G$ model, Fig. \ref{fig:TM},
which shows the transfer matrix spectrum, $E(k_x,k_y)$, for various
values of $\phi$.  Similarly to the Kitaev limit, minima in the
continuum of excitations are clearly identified at $(0,0)$,
$(\pi/3,2\pi/3)$, $(2\pi/3,-2\pi/3)$, and $(\pi,0)$. All, however, are
gapped. This can be understood in the context of Majorana fermions by
noting that the cylindrical geometry breaks the symmetry between $x$
bonds and $y,z$ bonds, which in turn can lead, for $\G>0$, to
anisotropic hopping amplitudes, and the gapping out of the
fermions. To corroborate this point, Fig. \ref{fig:E_per_bond} depicts
the energy density per bond as a function of $\phi$, displaying that
indeed the symmetry between bonds is broken for $\phi>0$.  

\begin{figure}[t]
  \centering
  \includegraphics[width=0.9\linewidth]{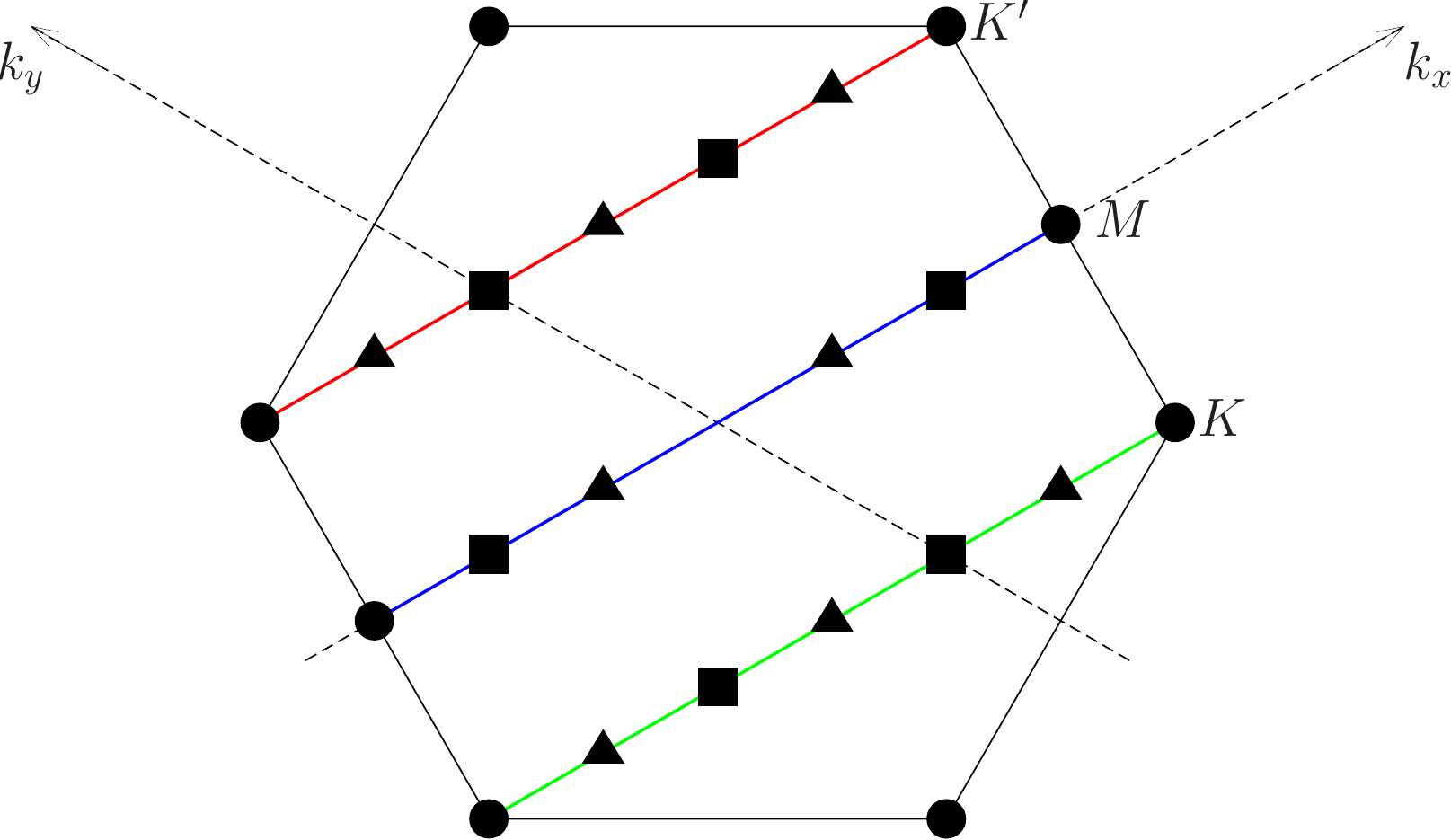}
  \caption{Allowed momentum cuts for the cylindrical geometry. Symbols
    indicate the positions of the soft two-particle excitations, as
    deduced from the TM spectrum. $\bullet$ indicate the leading soft
    modes at $K,K'$ and $M$ points. Additional soft modes are plotted
    for $\phi=0.03\pi\,(\blacktriangle)$ and
    $\phi=0.2\pi\,(\blacksquare)$. As $\phi$ is tuned between these
    values, we expect that these minima smoothly move from the squares
    ($\blacksquare$) to the triangles ($\blacktriangle$).}
\label{fig:bz}
\end{figure}

Several additional minima appear for $\phi>0$, with their momentum
position moving as $\phi$ is increased. Strikingly, these additional
minima seem to obey an underlying symmetry, i.e., a considerable
number of eigenvalues obey $E(k_x,k_y) = E(k_x + 2\pi/3,
k_y-2\pi/3)$. For instance, the $\phi=0.1\pi$ panel in
Fig. \ref{fig:TM} has a minimum near $(\pi/6,-2\pi/3)$ (green +),
which has a symmetric counterpart near $(5\pi/6,2\pi/3)$ (red x),
i.e., shifted in momentum by $(2\pi/3,-2\pi/3)$. An additional
counterpart is located near $(\pi/2,0)$ (blue circle), which can be
reached by inversion $\bk\to -\bk$, followed by the same shift in
momentum.  Interpreting the TM spectrum as being associated with
two-particle excitations, the above symmetry suggests the existence of
single-particle excitations which, in addition to inversion symmetry
$\varepsilon(-\bk) = \varepsilon(\bk)$, obey also
$\varepsilon(\bk)=\varepsilon(\bk\pm{\bf K})$, where $\pm{\bf K}$ are
the momenta at the Brillouin zone corners, $K$ and $K'$, respectively.
Figure \ref{fig:bz} shows the positions of the soft two-particle
excitations, for $\phi=0.03\pi$ and $\phi=0.2\pi$, further
demonstrating the above symmetry.

In summary, the features of the TM spectrum strongly indicate that the
paramagnetic phase of the $K-\G$ model harbours coherent excitations
commonly associated with quantum spin liquids. However, it is
difficult to determine the nature of this spin liquid phase, based on
the iDMRG data alone. On the one hand, the TM features suggest that in
the region $0<\phi<0.6\pi$ the $K-\G$ model harbours Majorana fermion
excitations, sharing basic properties with the ground state of the
ferromagnetic Kitaev model. On the other hand, the apparent transition
at $\phi=0.025\pi$ may indicate that there are two distinct spin
liquid phases with a sharp transition between them. In the next
section we introduce a mean-field approximation, which can be used to
elucidate the above results.

\section{Majorana mean-field theory}
\label{sec:MFT}

\subsection{Majorana spectrum}

Motivated by the iDMRG results of the previous section, we would like
to formulate a Fermionic mean-field theory which closely resembles the
exact solution of the Kitaev model. Therefore, following
Kitaev\cite{kitaev_anyons_2006}, we replace the spin operators in the
Hamiltonian with products of Majorana fermion operators, $2S_i^\a\to
ib_i^\a c_i$,
\begin{equation}
  \label{eq:Hc}
  \tilde H=-\sum_{\braket{ij}\a\b}K_{ij}^{\a\b}\,ib_i^\a b_j^\b\, ic_ic_j,
\end{equation}
where for $\braket{ij}$ a
$z$-type bond,
\begin{equation}
  \label{eq:Kdef}
  K_{ij}^{\a\b}=\frac{1}{4}\left\{
    \begin{array}{ccc}
      K &  & \a=\b=z \\ \G & & \a\ne\b\ne z \\ 0 & & \mbox{otherwise}
    \end{array}\right. .
\end{equation}
Similar definitions follow for $x$ and $y$-type bonds.  Here, the
Majorana fermion operators are normalized such that
$\{b_i^\a,b_j^\b\}=2\delta_{ij}\delta_{\a\b}$ and
$\{c_i,c_j\}=2\delta_{ij}$. The physical Hilbert space of the spin
Hamiltonian $H$ is then obtained by projecting the Majorana
Hamiltonian $\tilde H$ onto the subspace of states $\ket{\Psi}$ which
obey $D_i\ket{\Psi}\equiv
b_i^xb_i^yb_i^zc_i\ket{\Psi}=\ket{\Psi}$. Within a mean-field
approach, we can approximate $\tilde H$ with
\begin{equation}
  \label{eq:Hmf}
  H_{MF}\!=\!-\!\!\sum_{\braket{ij}\a\b}B_{ij}\,K_{ij}^{\a\b}\,ib_i^\a b_j^\b
  -\sum_{\braket{ij}}A_{ij}\,ic_ic_j+\sum_{\braket{ij}}A_{ij}B_{ij},
\end{equation}
where the fields $A_{ij}$ and $B_{ij}$ obey the mean-field
self-consistency equations on each bond,
\begin{eqnarray}
  \label{eq:AB}
  A_{ij} & = & \sum_{\a\b}K_{ij}^{\a\b}\braket{ib_i^\a b_j^\b}_B, \\
  \label{eq:BA}
  B_{ij} & = & \braket{ic_ic_j}_A.
\end{eqnarray}
Given the ground state $\ket{\Psi_0}_{MF}$ of $H_{MF}$, it is possible
to construct an approximate ground state for $H$ by projection onto
the physical Hilbert space, $\ket{\Psi_0}\approx \prod_i(1+D_i)/{2}\ket{\Psi_0}_{MF}$.
\begin{figure}[t]
  \centering
  \includegraphics[width=\linewidth]{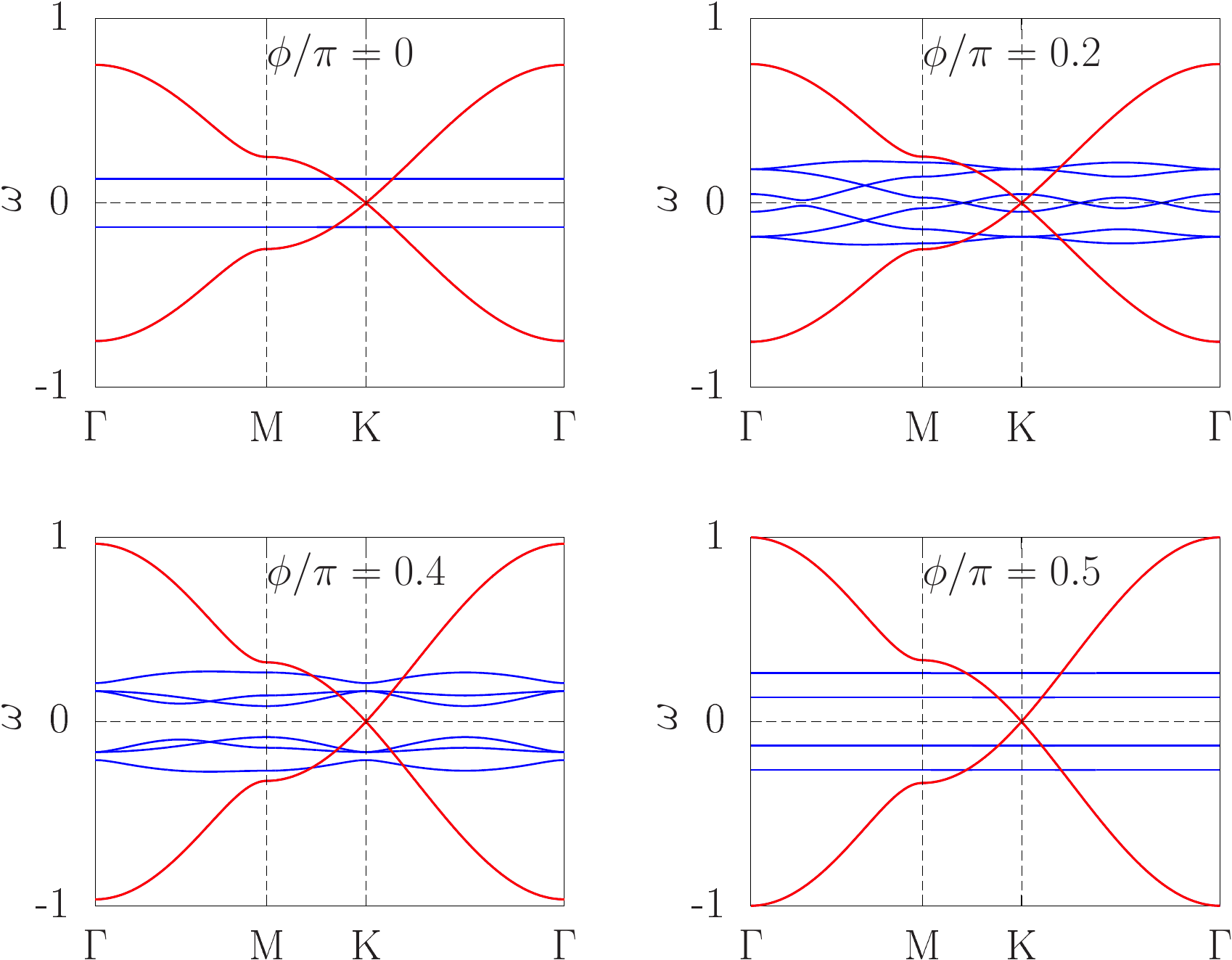}
  \caption{Band structure of $c$ (red) and $b$ (blue) Majorana
    fermions, plotted along high symmetry lines of the Brillouin zone,
    for several values of $\phi$ in the two dimensional thermodynamic
    limit.  The flat bands for $\phi=0$ are three-fold degenerate; for
    $\phi/\pi=0.5$, the lower energy flat bands are two-fold
    degenerate. When $\phi/\pi\sim 0.2$, there is a finite density of
    zero energy $b$ fermion states, around the $\rm K$ (and $\G$)
    points in the Brillouin zone.  At $\phi/\pi=0.4$ the $b$ fermion
    bands are still dispersive, but gapped.}
  \label{fig:Ek}
\end{figure}

It is straightforward to obtain a uniform, Z$_2$-flux-free, solution
of Equations (\ref{eq:AB}) and (\ref{eq:BA}), in the two-dimensional
thermodynamic limit. In the following we use the convention that in
$A_{ij},B_{ij}$ etc., the subscript $i$ indicates a site on the odd
sublattice and $j$ a site on the even sublattice.  Assuming that
$A_{ij}\equiv A$ on all bonds, we obtain $B_{ij}\equiv B = 0.5248$,
which is independent of $A$. Similarly, $A$ is independent of $B$, but
it does depend on the ratio $K/\G$. The mean-field ground state energy
per bond is given by $E_{MF} = -AB$. By solving $H_{MF}$, it is
possible to obtain the Majorana fermion spectrum, shown in
Fig. \ref{fig:Ek} along high symmetry lines in the Brillouin zone, for
several values of $\phi$. In the Kitaev limit, $\phi=0$, one finds a
single dispersing $c$-fermion band, with Dirac nodes at the K points
of the Brillouin zone, and whose band width is set by $A=K$. Three
flat bands describe the $b$-fermions, which are localized on the
bonds. For $\phi=\pi/2$, one obtains a similarly dispersing
$c$-fermion band, with band width $A=4\G/3$, and three flat bands
which describe the $b$-fermions localized on the hexagons. When both
$K$ and $\G$ are nonzero, the $b$-fermions are dispersive, and become
gapless in the parameter range $0.15 < \phi / \pi < 0.25$.
\begin{figure}[t]
  \centering
  \includegraphics[width=\linewidth]{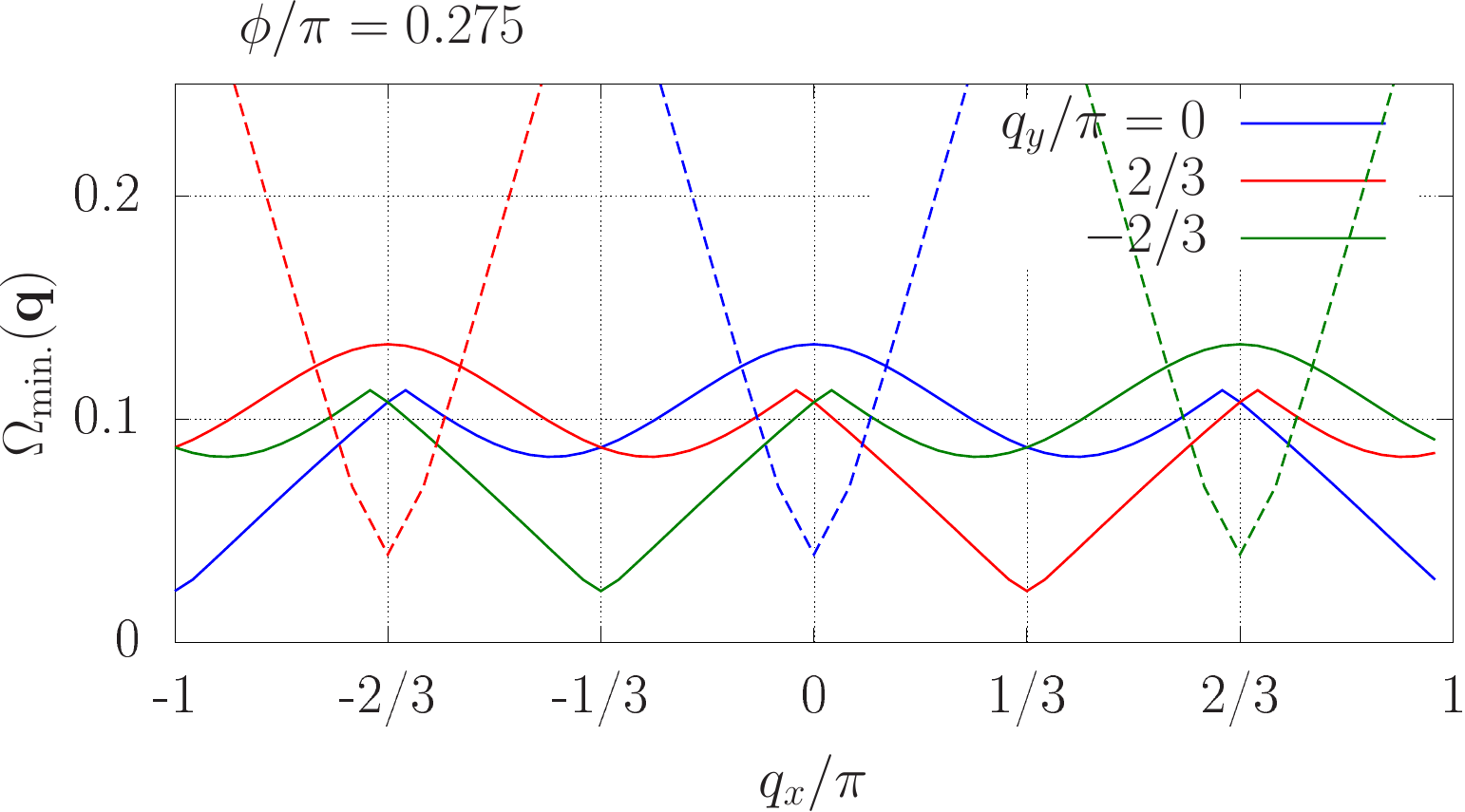}
  \caption{$b+c$ (solid) and $c+c$ (dashed) two-Majorana fermion
    spectrum plotted along the three momentum cuts allowed on a
    cylinder with a three unit cell circumference (see
    fig. \ref{fig:bz})}
  \label{fig:minex}
\end{figure}

\subsection{Two-Majorana spectrum}
\label{sec:MFT_2p}

We have suggested that the TM spectrum can be associated with
two-particle continua of fractionalized excitations, \emph{some of
  which} obey a symmetry relation
$E(k_x,k_y)=E(k_x+2\pi/3,k_y-2\pi/3)$. Interestingly, the Majorana
mean-field Hamiltonian, Eq. (\ref{eq:Hmf}), exhibits this symmetry for
the $b$ fermion spectrum, $\varepsilon_b(\bk)=\varepsilon_b
(\bk\pm{\bf K})$ (see Appendix \ref{sec:appMFT} for details), inducing
the same symmetry in the two-particle spectrum, $\Omega_{b+c}(\bq,\bk)
= |\varepsilon(\bq-\bk)|+|\varepsilon(\bk)|$, as well. We note that
this symmetry, $\Omega_{b+c}(\bq\pm{\bf K},\bk) =
\Omega_{b+c}(\bq,\bk)$, holds irrespectively of the properties of the
$c$-fermion spectrum. In Appendix \ref{sec:appiDMRG}, we show that
this is consistent with the iDMRG results, which exhibit this symmetry
in the TM spectrum even when the $K-\G$ coupling are anisotropic such
that the minima in the $c$ spectrum move away from the $K,K'$ points.

Thus, the form of $H_{MF}$ may give a good description of the
fractionalized excitations, as probed by iDMRG. However, due to the
strongly interacting nature of the $K-\G$ model, the actual amplitudes
$A_{ij}$ and $B_{ij}$ which should be used for such a description, as
well as the value of $\phi$ itself, will most likely be very different
from their values as determined by MFT. Nevertheless, we may still
compare the MF spectra with the TM spectra, demonstrating the
usefulness of $H_{MF}$. To do so we study the same cylindrical
geometry considered above using iDMRG.  As in the iDMRG calculation
where the cylindrical geometry breaks the symmetry between $x$ and
$y,z$ bonds, also here we choose different amplitudes $A_{ij}, B_{ij}$
for different bonds.  In Fig. \ref{fig:minex} we plot the minimal
energies required to excite two Majorana fermions, as given by
\begin{equation}
  \label{eq:minpole}
  \Omega_{\rm min.}(\bq) = \min_{\bk}\left(|\varepsilon_{b,c}(\bq-\bk)|+
   |\varepsilon_{c}(\bk)|\right),
\end{equation}
where $\varepsilon_{b,c}(\bk)$ is the Majorana spectrum of $H_{MF}$.
For finite anisotropy, $\varepsilon_{b,c}(\bk)$ opens a gap at all
allowed momenta, and consequently, also in $\Omega_{\rm min.}(\bq)$.
Nevertheless, the $K$ and $K'$ points remain soft, as in the TM
spectrum.  Furthermore, additional soft modes appear at finite $\G$ in
the $b+c$ spectrum, which obeys the symmetry $\Omega_{\rm
  min.}(\bq\pm{\bf K})=\Omega_{\rm min.}(\bq)$. For example, shifting
the solid green curve ($q_y=-2\pi/3$) in Fig. \ref{fig:minex} by
$q_x=2\pi/3$, yields the solid red curve ($q_y=2\pi/3$). By
demonstrating this symmetry, we conclude that $H_{MF}$ may give a good
description of the low energy excitations of the $K-\G$ model, as seen
in the TM spectrum. 

\section{Dynamic structure factor}
\label{sec:dynsf}

\subsection{Zero magnetic field}

\begin{figure}[t]
  \centering
  \includegraphics[width=\linewidth]{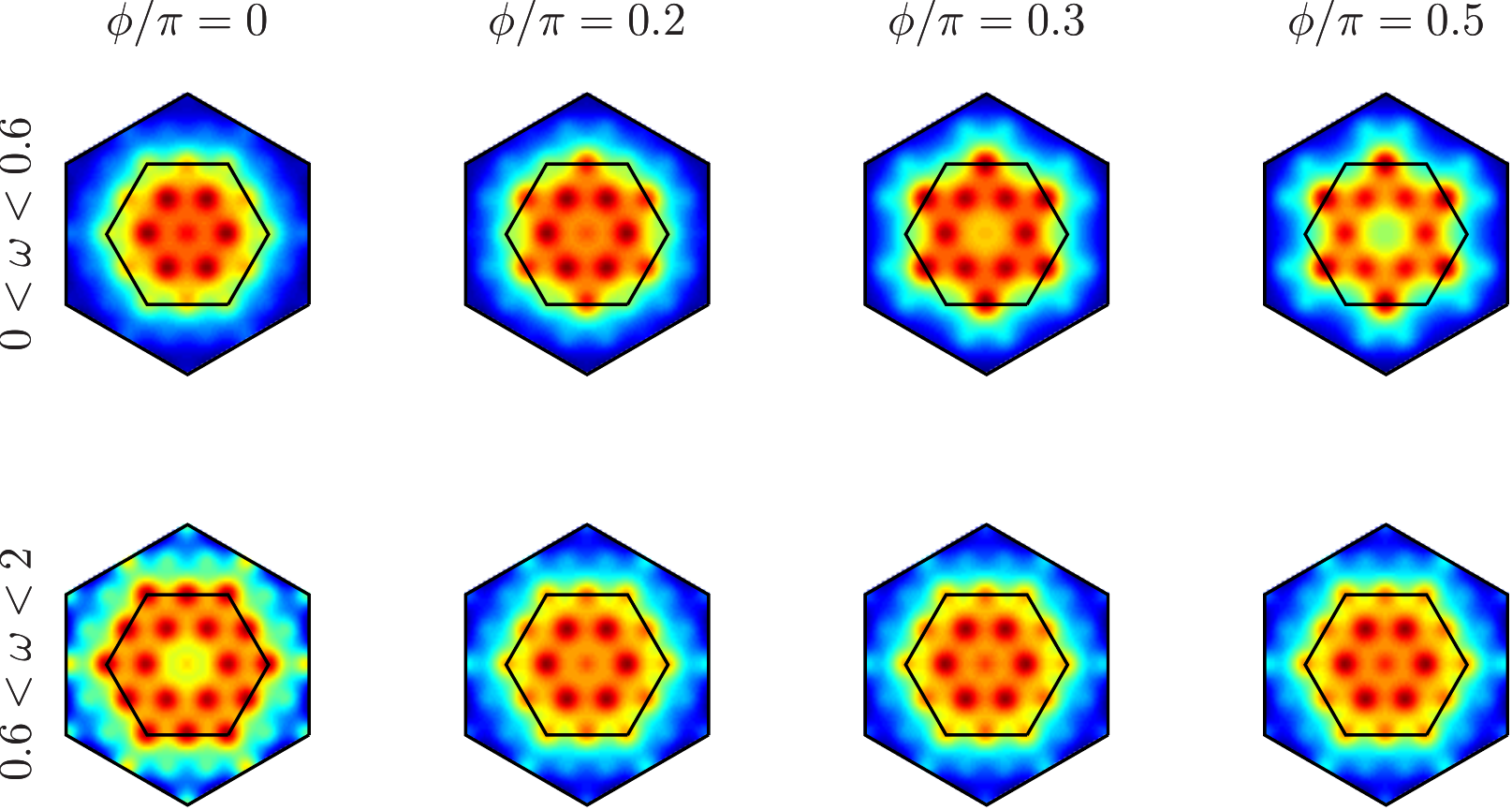} 
  \caption{Dynamical structure factor, $S(\bq,\omega)$, integrated
    over low (top) and high (bottom) frequencies, as obtained from
    exact diagonalization.  The discrete set of peaks in
    momentum space, resulting from the finite size of the studied
    cluster, was broadened in order to improve the visualization of
    the obtained pattern. The black lines depict the boundaries of the
    first and second Brillouin zones.}
  \label{fig:SED}
\end{figure}

Next, we turn to spectral signatures of the $K-\G$ spin liquid, which
can be observed in experiments. The dynamic structure factor, which is
probed in inelastic neutron scattering experiments, is defined as
\begin{equation}
  \label{eq:Sqw}
  S(\bq,\omega) = \sum_{j,\alpha}\int dt\braket{S_j^\alpha(t)
    S_0^\alpha(t=0)} e^{-i\bq\cdot(\br_j-\br_0)+i\omega t}.
\end{equation}
We calculated $S(\bq, \omega)$ using an ED method for a 24-site
cluster, see appendix \ref{sec:appDSF} for details. Fig. \ref{fig:EDA}
shows $S(\bq,\omega)$ for several values of $\phi$. The first evident
feature is the existence of a broad excitation continuum at high
frequencies. In the Kitaev limit, $\G=0$, most of the spectral weight
is found at relatively low energies, which is in agreement with literature
\cite{knolle_dynamics_2014,knolle_dynamics_2015}. As $\G$ is increased, the low
frequency spectral weight at the Brillouin zone center ($\G$ point) is
pushed towards higher frequencies, while a low $\omega$ signal remains
at the $\rm M$, $\rm Y$ and $\rm K/2$ (midway between $\rm K$ and
$\G$) points. In contrast to the spectra of the Kitaev-Heisenberg model
\cite{yamaji2016,gotfryd2017}, the spin gap at the Kitaev limit seems
to remain finite even in the presence of large $\Gamma$.
The difference in momentum dependence between high and
low frequencies is clearly evident by integrating over different
ranges of $\omega$.
\begin{figure}[t]
  \centering
  \includegraphics[width=\linewidth]{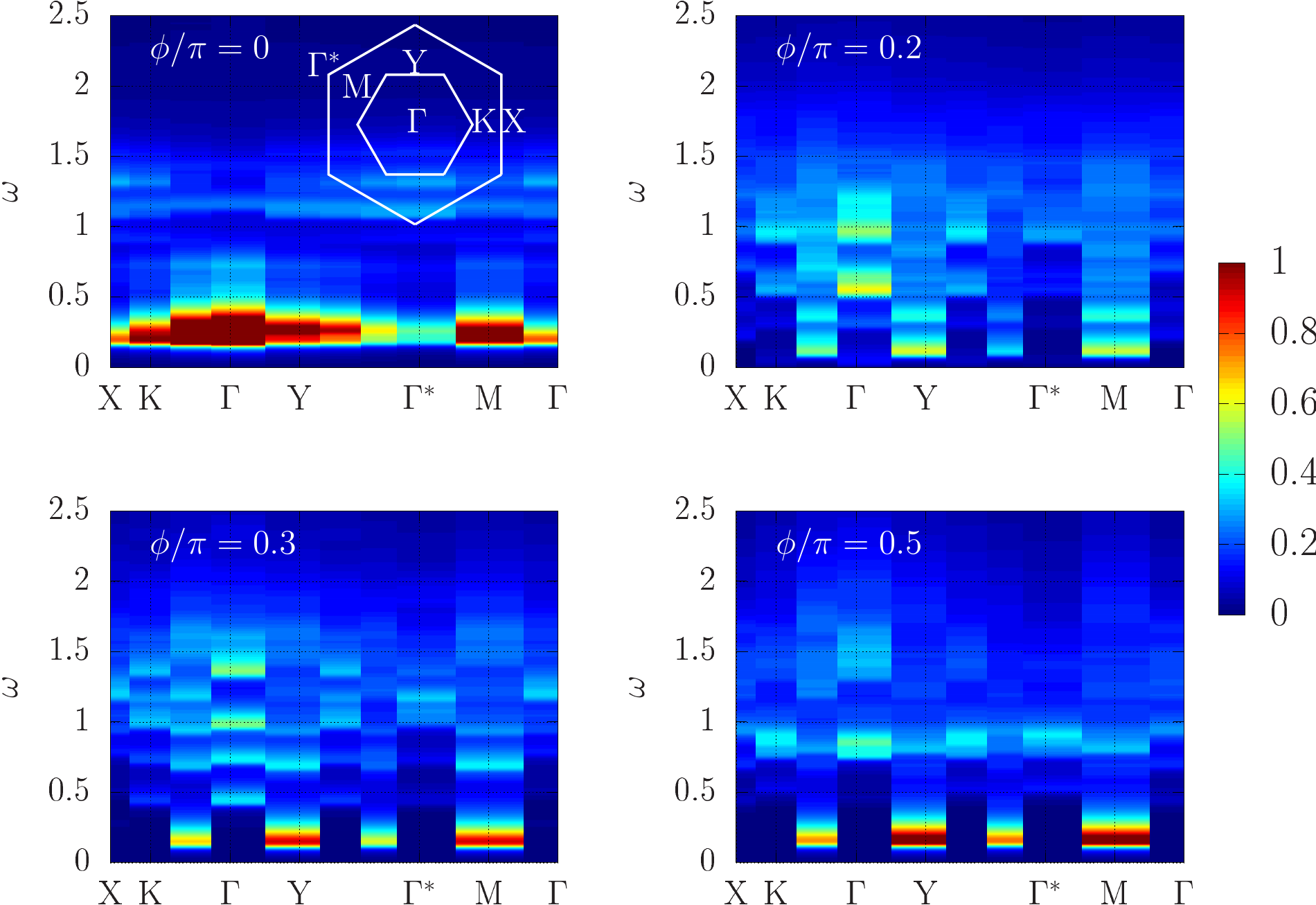}
  \caption{Intensity plot of the dynamic structure factor, $S(\bq,
    \omega)$, presented along high symmetry lines of the Brillouin
    zone for several values of $\phi$, as obtained using the ED
    method. Pseudo-color indicates relative intensity. Inset:
    Brillouin zone with labels of symmetry points used in this
    figure.}
  \label{fig:EDA}
\end{figure}
In Fig. \ref{fig:SED} we show the \emph{relative} intensity of
$S(\bq,\omega)$, integrated over low and high frequency ranges.  In
the Kitaev limit, $\phi=0$, $S(\bq,\omega)$ is rather featureless. As
$\G$ is increased, $S(\bq,\omega)$, integrated over a low frequency
range, shows a star shaped pattern, similar to the pattern seen in the
$\a$-RuCl$_3$ neutron scattering experiments at low energies. In
contrast, integrating over a range of higher frequencies, shows an
almost featureless momentum dependence even for finite $\G$, again, in
qualitative agreement with the experiments.
\begin{figure}[t]
  \centering
  \includegraphics[width=\linewidth]{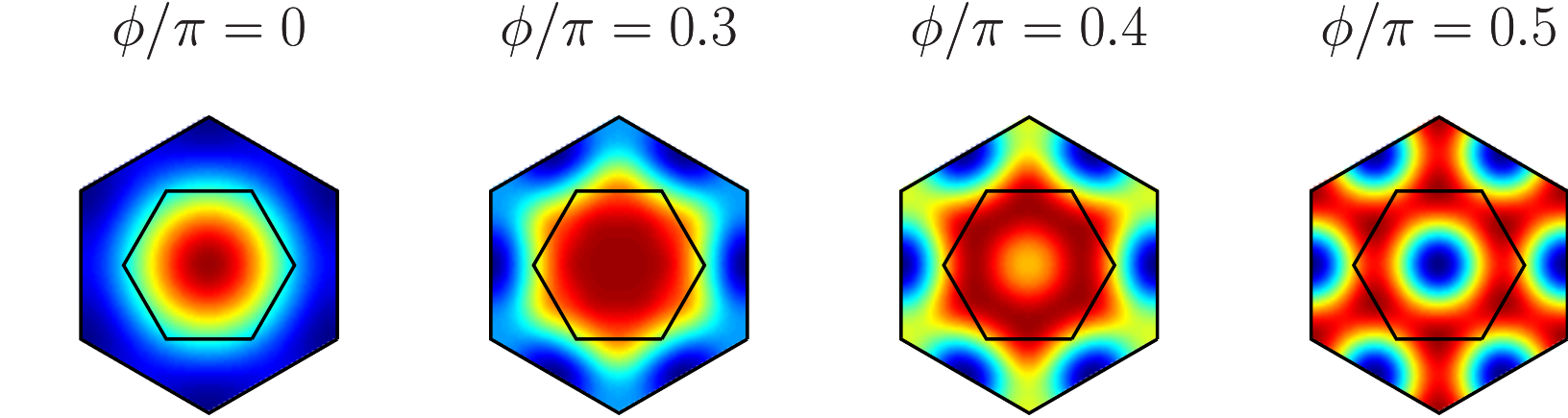}
  \caption{Equal-time structure factor $S(\bq)$, obtained using the
    Majorana mean-field theory.}
  \label{fig:Smf}
\end{figure}

To calculate the dynamic spin structure factor in the context of the
Majorana MFT, one must consider Z$_2$ flux excitations with respect to
the ground state, since each spin operator inserts a
flux\cite{baskaran_exact_2007, knolle_dynamics_2014}. Technically,
this requires solutions to Eqns. (\ref{eq:AB}) and (\ref{eq:BA}) which
go beyond the uniform ansatz considered here. It is however still
possible to approximately calculate the equal-time spin structure factor,
\begin{eqnarray}
  \label{eq:Sq}
  S(\bq) & = & \sum_{i,\alpha}\braket{S_i^\alpha S_0^\alpha}
  e^{-i\bq\cdot(\br_i-\br_0)} \nonumber \\ & \approx & 
  \frac{1}{4}\sum_{i,\alpha}\braket{b_i^\a b_0^\a}_B
  \braket{c_ic_0}_Ae^{-i\bq\cdot(\br_i-\br_0)}.
\end{eqnarray}
Since $S(\bq)=\int (d\omega/2\pi)\, S(\bq,\omega)$, and noting that
according to the ED results, most of the dynamic structure factor
signal is concentrated at low frequencies, we expect that $S(\bq)$
resemble the integrated low frequency patterns of
$S(\bq,\omega)$. Indeed, as seen in Fig. \ref{fig:Smf}, the
mean-field theory reproduces the star shaped pattern seen both in
experiments and in the ED calculation.

\begin{figure}[t]
  \centering
  \includegraphics[width=\linewidth]{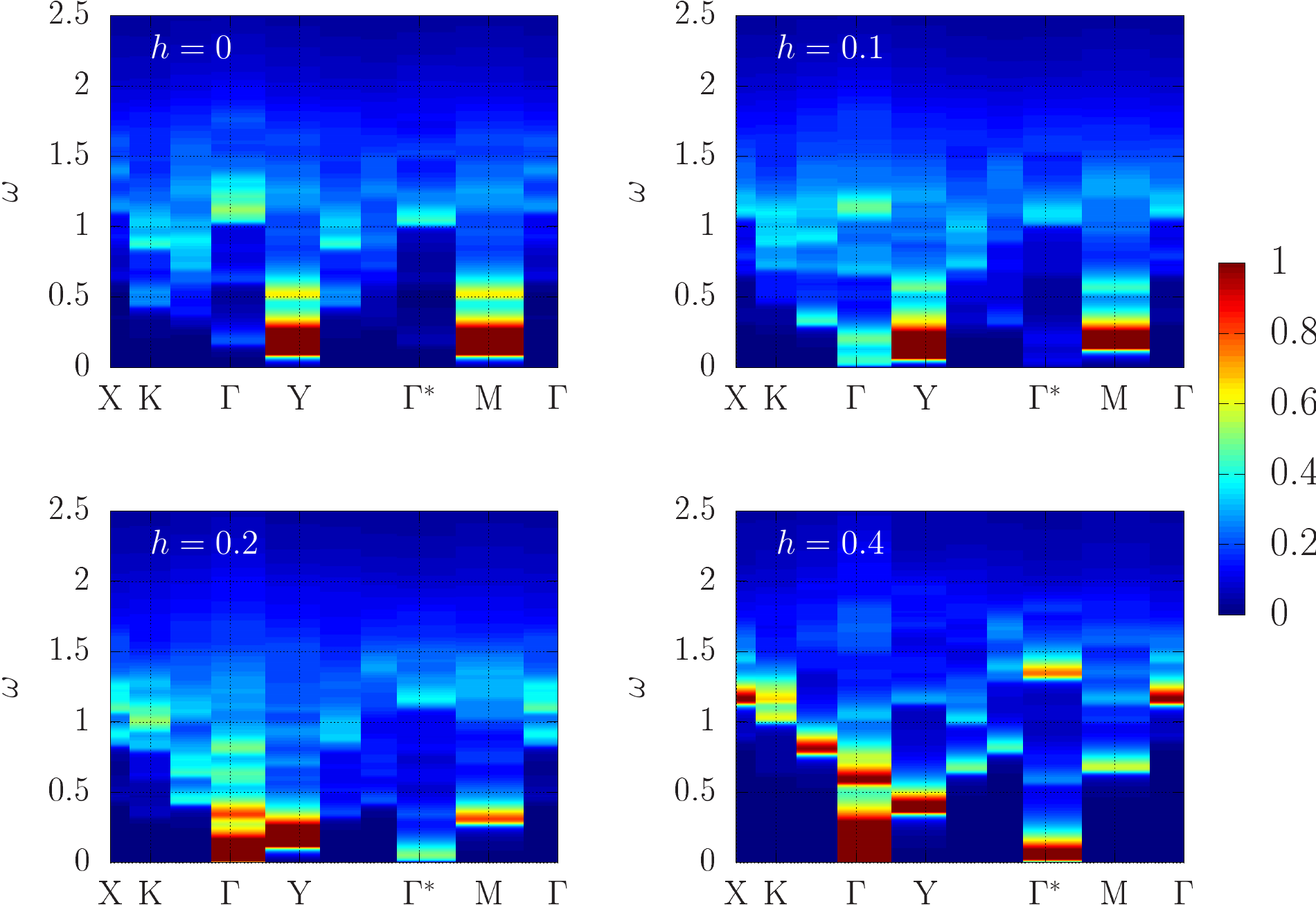}
  \caption{Dynamic structure factor for $\phi/\pi=0.2$, $J_3=0.05$,
    and several values of the magnetic field $h_{\perp c^*}$. Symmetry
    points labeled as in Fig. \ref{fig:EDA}.}
  \label{fig:EDB}
\end{figure}

\begin{figure}[t]
  \centering
  \includegraphics[width=\linewidth]{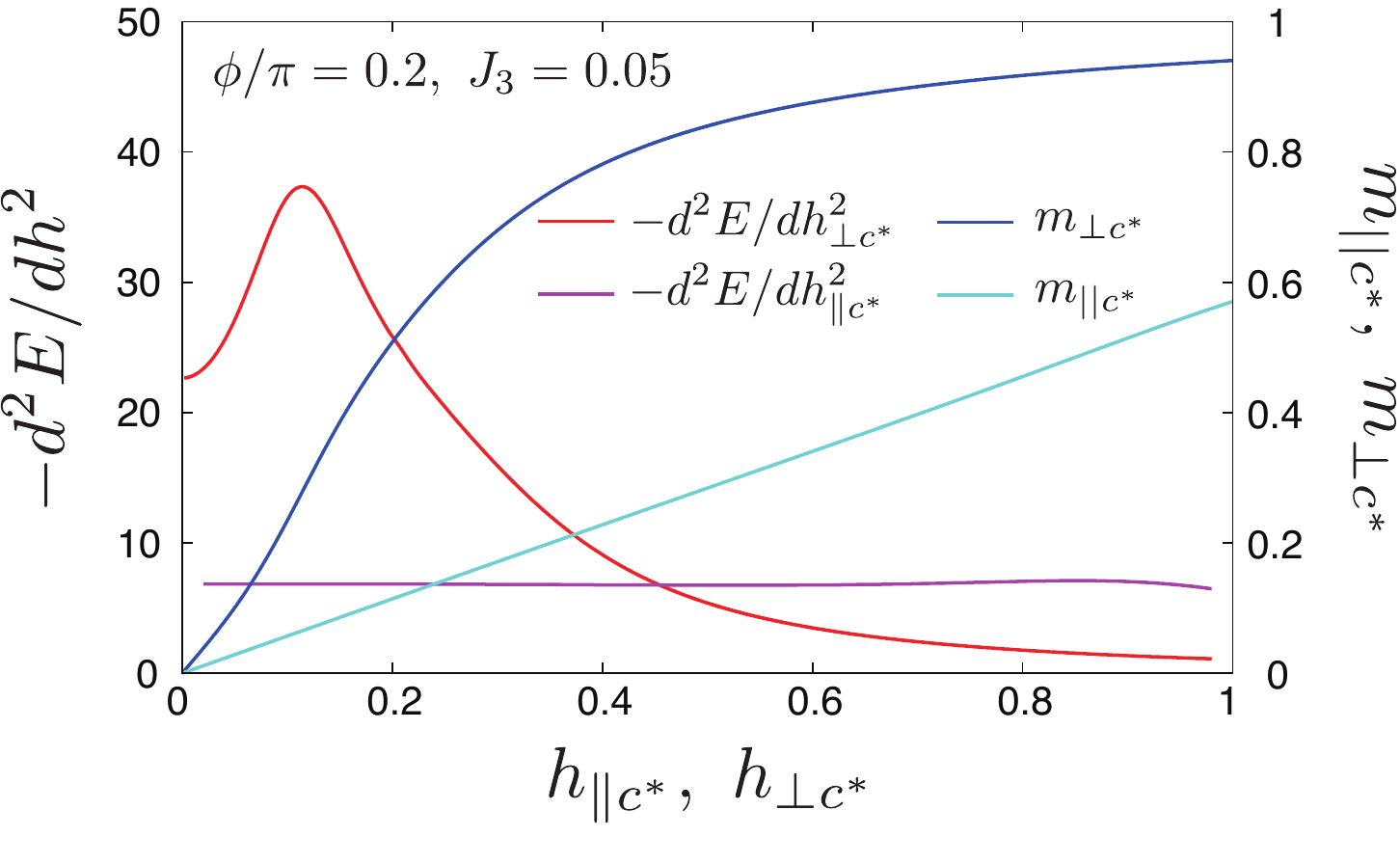}
  \caption{Magnetization and second derivative of the energy as a
    function of magnetic field, parallel , $h_{||c^*}$, and
    perpendicular, $h_{\perp c^*}$, to $c^* = (1,1,1)$.} 
  \label{fig:MED}
\end{figure}

\subsection{$J_3$ terms and finite magnetic field}

Zig-zag magnetic ordering, similar to the magnetically ordered state
observed in $\a$-RuCl$_3$ at low temperatures, can be stabilized by
adding a third neighbor Heisenberg term,
$J_3\sum_{\a,\braket{ij}\in{\rm3^{rd}n.n.}}S_i^\a S_j^\a$ to the
Hamiltonian, Eq. (\ref{eq:Hb}).  At this level, nearest neighbor
Heisenberg terms are less important since microscopic
calculations\cite{winter_challenges_2016} suggest that they are
smaller, and $J_3$ is enough to stabilize the zig-zag magnetic order.
Furthermore, it is possible to suppress this ordering tendency by
applying a magnetic field, $-\sum_{i\a} h^\a S_i^\a$. This is evident
in Fig \ref{fig:EDB}, which displays $S(\bq,\omega)$ for
$\phi/\pi=0.2$, $J_3=0.05$ and several values of the in-plane magnetic
field $h_{\perp c^*}\propto(-1,1,0)$, perpendicular to $c^* =
(1,1,1)$, which corresponds to the out-of-plane direction. With
$h_{\perp c^*}=0$ the low $\omega$ spectral weight is increased at the
M and Y points, but not at ${\rm K}/2$. Notice, however, that most of
the zone center ($\G$ point) spectral weight is found at relatively
high energies\cite{winter_breakdown_2017}.  This aspect of the spectra
is similar to the case with $J_3=0$, $h=0$, shown in
Fig. \ref{fig:EDA}.  When $h_{\perp c^*}$ is increased beyond
$h_{\perp c^*}\sim 0.1$, the zone center spectral weight shifts
towards lower energies, and a continuum of excitations emerges.

Fig.  \ref{fig:MED} shows magnetization curves as obtained with ED,
for magnetic fields pointing parallel, $h_{||c^*}$, and perpendicular,
$h_{\perp c^*}$, to $c^*$. A peak at $h_{\perp c^*}\sim 0.1$ in the
second derivative of the energy $-d^2E/dh_{\perp c^*}^2$, indicates an
apparent transition away from zig-zag order.  In addition, the
magnetization curves display an easy axis anisotropy, also consistent
with $\a$-RuCl$_3$ experiments\cite{johnson_monoclinic_2015,
  sears_magnetic_2015, vojta2017}.  A simple mean-field analysis qualitatively
explains the easy-plane anisotropy as follows: If a ferromagnetically
ordered moment $\vec{m}$ is assumed as a classical Weiss field, the
mean-field energy $E_{\rm mf}$ is obtained as
\begin{eqnarray}
  E_{\rm mf}/N &=& -\frac{K-3J_3}{2} (m_x^2+m_y^2+m_z^2)
  \nonumber\\
  &&+\Gamma (m_y m_z + m_z m_x + m_x m_y)
  \nonumber\\
  &=& -\frac{K+\Gamma-3J_3}{2}(m_x^2+m_y^2+m_z^2)\nonumber\\
  &&+\frac{\Gamma}{2}(m_x + m_y + m_z )^2, 
\end{eqnarray}
which is minimized, for finite $\G>0$, when $m_x + m_y + m_z=0$ is
satisfied, i.e., when the magnetic moments are in-plane.

\section{Conclusions}

In this work, we investigated a spin model with both the Kitaev ($K$)
and symmetric-anisotropic ($\Gamma$) interactions on the honeycomb
lattice using iDMRG, exact diagonalization, and Majorana mean-field
theory. This model is strongly motivated by recent experiments on
$\alpha$-RuCl$_3$, where $K$ and $\Gamma$ are likely to be the
dominant exchange interactions.

We found strong numerical evidence for the existence of a quantum spin
liquid for arbitrary ratio of $\Gamma/K$ for ferro-like Kitaev
interactions in iDMRG. In particular, the entanglement entropy remains
very high in this entire region while we do not see any sign of
magnetic order in iDMRG computations. In contrast, we found a magnetically
ordered state with very small entanglement entropy on the antiferro-like
Kitaev side. Moreover, we demonstrated the existence of coherent
two-dimensional multi-particle excitations using the correspondence
between transfer-matrix eigenvalues and the lower boundary of
multi-particle excitation spectrum. The cylinder geometry in iDMRG
induces an anisotropy in bond-dependent energy, which is expected to
move the locations in momentum space of low energy excitations. We
show that this can indeed be seen in the transfer-matrix spectra. The
existence of such two-dimensional coherence excitations without
magnetic order is a very strong evidence of quantum spin liquid.

In order to make direct connection to neutron scattering experiments,
we computed the dynamical structure factor for the $K-\Gamma$ model
without and with a small third neighbor Heisenberg interaction $J_3$
using exact diagonalization of 24-site cluster. The $J_3$ on top of
the $K-\Gamma$ interactions is shown to drive a transition to the
zig-zag order in agreement with previous numerical
studies\cite{winter_challenges_2016, winter_breakdown_2017,
  catuneanu_realizing_2017}. Upon introduction of $\Gamma$ starting
from the ferro-like Kitaev interaction, the dynamical structure factor
develops the scattering continuum with star-like intensity profile at
low energies, just like what is seen in recent neutron scattering
experiment\cite{banerjee_neutron_2017}. The magnetic field dependence
of the dynamical structure factor is also investigated when $J_3$ is
finite such that the ground state is the zig-zag magnetic order in
zero field. There is a transition to a paramagnetic state with
dominant scattering intensity at the zone center when the external
magnetic field along the honeycomb plane reaches about $1/10$ of the
largest exchange interactions, namely $K$ or $\Gamma$. This is seen in
the magnetization profile and dynamical structure factor computed in
exact diagonalization. All of these features are consistent with
recent experimental data.

Further we used the Majorana mean-field theory to gain analytical
insight in these numerical results. For example, we computed single
and two-particle excitation spectra for the $K-\Gamma$ model showed
they exhibit an emergent symmetry in their momentum dependence, which
is also found in the transfer matrix spectrum.  The equal-time
structure factor and real-space spin correlations computed in the
Majorana mean-field theory are also consistent with main features in
the exact diagonalization results.

Combining all these results together, we conclude that the mysterious
scattering continuum seen in the neutron scattering experiment on
$\alpha$-RuCl$_3$ may come from a nearby quantum spin liquid supported
by $K-\Gamma$ interactions. In our numerical computations, the spin
liquid phases at finite $\Gamma/K$ show qualitatively the same
behavior as the Kitaev spin liquid. We do see, however, a jump in the
bond-dependent energy at some value of $\Gamma/K$ in iDMRG on cylinder
geometry, which causes a small kink in the entanglement entropy. This
can be interpreted as a “meta-nematic” transition, where the
bond-anisotropy (or broken 3-fold rotation symmetry) increases
abruptly. Whether such a transition would survive in the 2D limit is
not clear to us at present.  If it does survive, we would need to
consider two possible scenarios.  (i) The transfer matrix spectra on
both sides of the transition share some qualitative features,
suggesting that they are actually the same spin liquid phase, while
the apparent transition may be interpreted as a Lifshitz transition on
the Fermi surface topology of the underlying quasiparticles. (ii)
Although the transfer matrix spectra may be described using Majorana
fermions both before and after the transition, the underlying spin
liquid ground states may be distinct.  These questions will have to be
addressed in future theoretical investigations. Further experimental
data in external magnetic field would provide additional clues for the
validity of the assumption that the $K-\Gamma$ or $K-\Gamma-J_3$ is a
good minimal model for $\alpha$-RuCl$_3$.

\section*{Acknowledgements}

We would like to thank Andrei Catuneanu, Yin-Chen He, Masatoshi Imada,
Hae-Young Kee, Young-June Kim, Stephen Nagler, Tsuyoshi Okubo, Natalia
Perkins, Roser Valenti and Ruben Verresen for useful discussions. YY further thanks
Mitsuaki Kawamura for his technical support.  YY was supported by
PRESTO, JST (JPMJPR15NF).  The ED computation has partly been done
using the facilities of the Supercomputer Center, the Institute for
Solid State Physics, the University of Tokyo.  MG and FP acknowledge
support from the German Research Foundation (DFG) via SFB 1143 and
Research Unit FOR 1807.  GW and YBK are supported by the NSERC of
Canada, Canadian Institute for Advanced Research, and the Center for
Quantum Materials at the University of Toronto.

\appendix
\section{iDMRG and Transfer Matrix}
\label{sec:appiDMRG}

This appendix is devoted to the infinite Density Matrix Renormalization
(iDMRG) method and the transfer matrix spectrum.  The first part
exposes the geometry used in order to apply iDMRG onto a 2D lattice
model.  We present more in-depth discussion of possible
finite size effects due to the finite circumference.  The second part
introduces the transfer matrix spectrum and its connection to the 
excitation spectrum.

\emph{Infinite Density Matrix Renormalization Group.} We use iDMRG
\cite{white_density_1992,mcculloch_infinite_2008,kjall_phase_2013} to
study ground state properties of the K-$\Gamma$ model.  Initially
developed for 1D systems, it has been successfully applied on 2D systems
by wrapping the lattice on a cylinder and mapping the cylinder to a
chain.  Furthermore employing translational invariance enables to study
infinite cylinders\cite{mcculloch_infinite_2008,kjall_phase_2013}.
Using the cylinder geometry, one dimension of the lattice is finite
and leads to a discretization of the related reciprocal vector.  Thus,
accessible momenta lie on lines in reciprocal space.  We chose
cylinder geometries, i.e. circumference $L_{circ}$ and unit cell, such
that the accessible momentum lines go through the gapless nodes 
of the isotropic Kitaev spin liquid, that are located at the $K$-points 
of the first Brillouin zone.  The results presented here and in the main
text are obtained using a rhombic unit cell and a narrow cylinder with
$L_{circ} = 6$ sites circumference.

We extend the discussion of the main text by considering the average
energy $\langle E_{x,y,z} \rangle$ of the $x,y$ and $z$ bonds for $0
\le \phi \le 0.5 \pi$, which is presented in
Fig. \ref{fig:E_per_bond}.  In the limit of small $\Gamma$
almost no anisotropy exists, which indicates negligible finite size
effects caused by the cylinder geometry.  Once $\Gamma$
increases, the anisotropy raises and reaches $\max(\langle
E_{y,z}\rangle/\langle E_x \rangle) \approx 2$ near $\phi = 0.3 \pi$.
Using wider cylinders with $L_{circ} = 12$ reduces the anisotropy only
slightly.  This suggests, that a $\Gamma$-like interaction is highly
sensitive to finite size and a cylinder geometry and presumably causing
the gapless Dirac nodes to shift away from the K-points in reciprocal
space.

\emph{Transfer Matrix Spectrum.} The transfer matrix (TM) of a 
wave function encoded as an
infinite matrix product state (iMPS) contains full information about
the static correlations \cite{zauner_transfer_2015}.  Intuitively, the TM translates the
iMPS by a lattice vector along the chain in 1D or the cylinder in 2D.
For Hamiltonians with only local interactions, the static 
correlations are related to the spectral
gap \cite{hastings_locality_2004}, e.g. $\xi\sim 1/\Delta$ for $z=1$.
This statement has been extended by Zauner et
al. \cite{zauner_transfer_2015} to also include momentum, such that
the length scale of the decay of static correlations with a
momentum $\bm k$ gives an upper bound on the spectral gap
$\epsilon(\bm \tilde k)$ at $\bm \tilde k$ close to $\bm k$.  Hence, 
the TM spectrum, a ground state property, provides
information about the position of the minimal energy excitation within 
the reciprocal space. 
A connection between the TM eigenvalues and the exact excitation energies can 
only be made knowing the Lieb-Robinson velocity \cite{lieb_finite_1972}, e.g., from dynamics.
Here, the Lieb-Robinson velocity is not known and thus the quasi-energies $E_i = -\log \lambda_i$, where $\lambda_i$ are the eigenvalues of the TM, are \emph{only} given up to an overall energy scale of the Hamiltonian.

On the cylinder geometry and If the symmetry upon translation along the cylinder's 
circumference is not broken, the transverse momentum $k_y$ is a good quantum number. 
Then, for each $k_y$ independently a set of $\lambda_i$ exists with a longitudinal momentum $k_x = \arg \lambda_i$ corresponding to the momentum of minimal energy excitation. 

\begin{figure}
	\includegraphics[width=\linewidth]{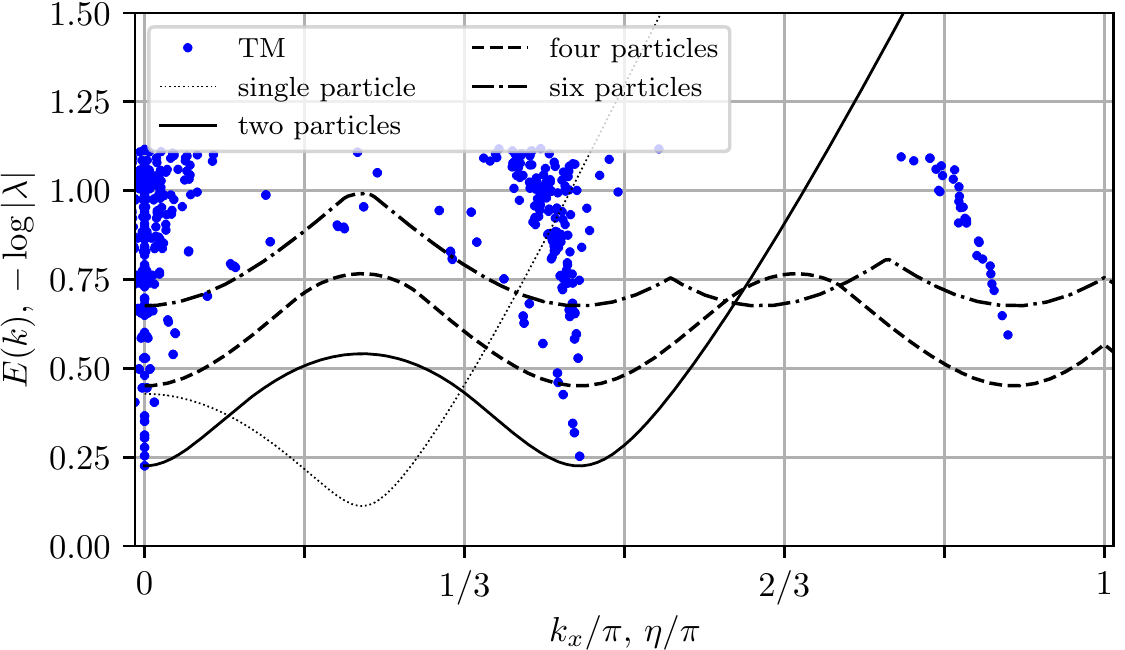}
	\caption{Quasi-energies $E(k) = -\log \lambda_i$ of the \emph{regular transfer matrix} spectrum $\lambda_i$ compared to analytical excitation energies for single- (thin dotted), two- (solid), four- (dashed), and six-particle excitations (dash-dotted) of the anisotropic XY-Heisenberg chain with transverse field. The parameters are: $\gamma = 0.1$, $h=0.75$. The analytical results are scaled by a factor $a=1.73$ in order to match the $\min(E_i)$ with $\min (E_{\text{2p}}(k))$, where $E_{\text{2p}}(k)
$ is the lower edge of the two-particle excitation band.} 
	\label{fig:XX_TM_g01_h075}
\end{figure}

In Fig. \ref{fig:XX_TM_g01_h075} we illustrate an exemplary TM spectrum for the anisotropic XY-Heisenberg chain with transverse field $h$ and anisotropy $\gamma$:
\begin{equation}
	H= -J \sum_{i} \left[ (1+\gamma) S^x_i S^x_{i+1} + (1-\gamma) S^y_i S^y_{i+1} + h S^z_i \right]   ~,
\end{equation}
This model can be solved exactly \cite{lieb_two_1961,katsura_statistical_1962,barouch_statistical_1970,barouch_statistical_1971, schmidt_excitation_1974} and its energy spectrum is known \cite{schmidt_excitation_1974}.
The plot compares the TM spectrum $E_i(k_i) = -\log \lambda_i$ (blue dots) with the analytical single- and multi-particle excitations (lines).
The position $k_i = \arg(\lambda_i)$ of the minimal eigenvalues $E_i$ coincide nicely with the minimum in the excitation bands with an even number of particles.
Single particle excitations are not present in the \emph{regular} TM.

\begin{figure} 
	\includegraphics[width=\linewidth]{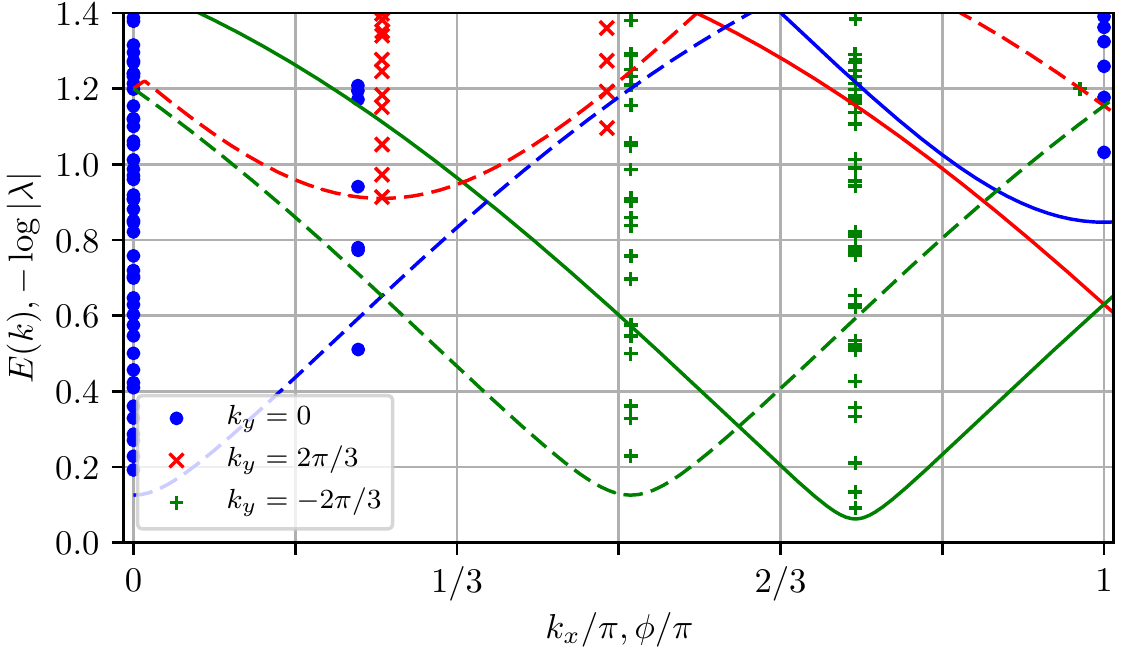}
	\caption{Quasi-energies $E(k)$ of the anisotropic Kitaev model with $K_x=-1$, $K_y=-1.2$, and $K_z=-0.9$. Lines display the analytic results of the single-particle (solid) and two-particles (dashed) lowest energy excitation \cite{kitaev_anyons_2006} on the corresponding momentum cuts: $k_y = 0$ (blue), $k_y = 2\pi/3$ (red), and $k_y = -2\pi/3$ (green). The analytic results are scaled by a factor $a = 0.38$.}
	\label{fig:TM_aniso_FMKSL}
\end{figure}

We provide a second example related to the model investigated in the main text.
Taking the limit $\phi \rightarrow 0$ of eq. \ref{eq:Hb} one obtains the exactly solvable Kitaev model on a honeycomb lattice \cite{kitaev_anyons_2006}.
If the Kitaev couplings are isotropic $K_x = K_y = K_z$, the model exhibits gapless Dirac nodes at the $K$-points of the Brillouin zone.  
As long as $|K_\alpha| < |K_\beta| + |K_\gamma|$ with $\alpha, \beta, \gamma = \{x,y,z\}$, the excitation remain gapless.
Once the $K$'s are tuned away from isotropy, the Dirac node moves and odd-numbered particle excitations get separated in reciprocal-space from even-numbered. 
Furthermore, the nodes may leave the allowed momenta cuts of the cylinder geometry introducing an effective gap. 
In Fig. \ref{fig:TM_aniso_FMKSL} we present a comparison of the TM spectrum with analytical results for the single- and two-particle spectrum clearly illustrating a correspondence between $(k_x, k_y)$ of the TM eigenvalues and the minimum of the excitations bands. 
Remarkably, the TM spectrum recovers single-, three-, etc. particle excitations as well as two-, four-, etc. particle excitations.

The reader will find a more rigorous and detailed explanation as well as more examples in Zauner et al.\cite{zauner_transfer_2015}.

\begin{figure}
  \includegraphics{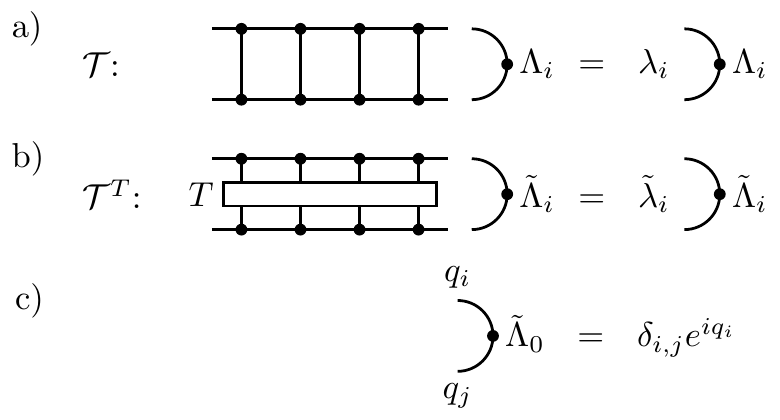}
  \caption{Schematic representation of a) the \emph{regular} and 
    b)~the \emph{mixed} transfer matrix $\mathcal{T}$ with translation
    $T$ applied along the circumference. c) Dominant eigenvector $\tilde \Lambda_0$ of
    $\mathcal{T}^T$ determines the $q$ quantum numbers associated with
    each bond leg.}
  \label{fig:TMillustration}
\end{figure}

\emph{Implementation.} We turn now to the technical realization of 
obtaining the momentum resolved TM spectrum.
Let $\lambda_i$ be the eigenvalues of the transfer matrix with the
ordering $|\lambda_0| > |\lambda_1| \ge |\lambda_2| \ge ...$ .  By
definition the dominant eigenvalue is $|\lambda_0| = 1$.  Generally,
$\lambda_i$ are complex and can be decomposed as $\lambda_i =
|\lambda_i| e^{i\eta}$.  The angle $\eta$ is connected to the momentum
$k_x$ along the chain or cylinder.  Exploiting the rotational
invariance of the Hamiltonian on the cylinder geometry yields the
transverse momentum $k_y$ as will be explained now.  In the following,
we require the iMPS to be in \emph{canonical
  form}\cite{vidal_efficient_2003}.  A translation with a lattice
vector along the circumference keeps the Hamiltonian invariant and as
such $k_y$ can be treated as a regular quantum number.  We extract
$k_y$ by computing the dominant eigenvector $\tilde \Lambda_0$ of the
\emph{mixed} transfer matrix constructed out of the ground state iMPS
and a iMPS with the translation applied, see also
Fig. \ref{fig:TMillustration}b).  We like to remark, that the
translation along the circumference is simply given by a permutation
of sites within a ring.  If the iMPS is sufficiently converged and the
applied translation is indeed a symmetry, then the dominant eigenvalue
$\tilde \lambda_0$ of the \emph{mixed} TM is $1$.  Its eigenvector
$\tilde \Lambda_0$ has a diagonal form with eigenvalues $|\tilde
\lambda_i| e^{i q}$ and $q$ being discretized in steps $2\pi/n$, where
$n$ is the number of unit cells around the cylinder.  If Schmidt
values are degenerate, the diagonal form becomes block diagonal with
blocks for each set of degenerate Schmidt values.  Each block can be
diagonalized separately by a unitary transformation which is then
applied to the non-translated iMPS.  The momentum quantum number $q_i$
are associated with the entries $i$ along a bond leg in the same way
as Schmidt values are.  The TM connects states $i$ and $j$ with
corresponding $q_i$ and $q_j$, hence the transverse momentum is given
by $k_{y,(i,j)} = q_j-q_i$.  The $k_y$ label of $\lambda_i$ can be
read off from its eigenvector $\Lambda_i$ due to the fact, that
$\Lambda_i$ has only non-zero entries with the same change of the
quantum number $q_i-q_j$.

\emph{Anisotropic $K-\G$ model.} 
Here, we provide TM spectra in the case of anisotropic $K-\G$ couplings in order to strengthen the discussion in the main text, section \ref{sec:TM_KG} and \ref{sec:MFT_2p}, about the symmetry of the b-fermions.

\begin{figure} 
	\includegraphics[width=\linewidth]{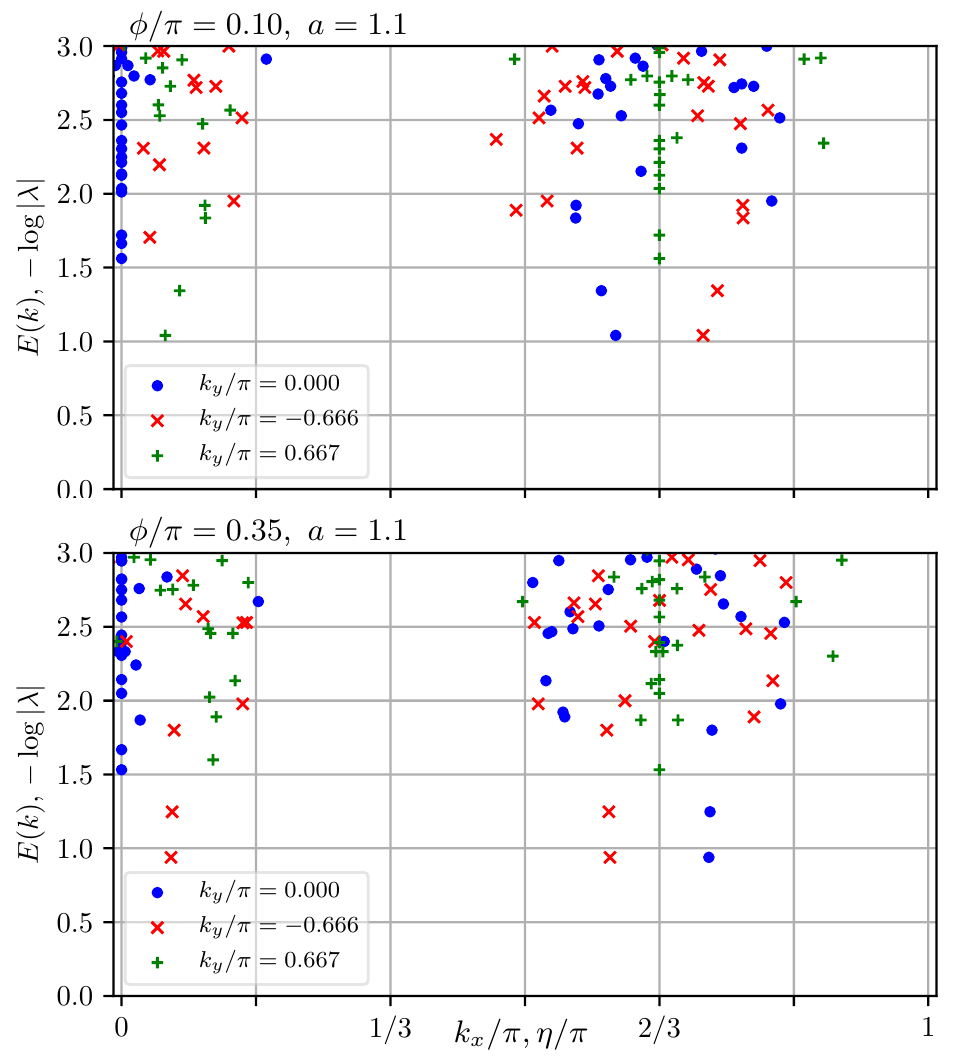}
	\caption{
	Quasi-energies $E(k)$ of the anisotropic K$\Gamma$ model for 
	$\phi = 0.1$ (top) and $\phi=0.35$ (bottom) with an anisotropy $a=1.1$ according to eq. \ref{eq:TM_anisoKG}.}
	\label{fig:TM_aniso_p010p035a11}
\end{figure}

We introduce an anisotropy by scaling the coupling parameter $K_\alpha$,$\G_\alpha$ by a factor $a$
\begin{equation}
	\begin{aligned}
		(K_x, \G_x) &\rightarrow (a K_x, a \G_x) \\
		(K_y, \G_y) &\rightarrow ((3/2 - a/2) K_y, (3/2 - a/2) \G_y) \\
		(K_z, \G_z) &\rightarrow ((3/2 - a/2) K_z, (3/2 - a/2) \G_z)~,
	\end{aligned}
	\label{eq:TM_anisoKG}
\end{equation} 
with respect to the x-bond. The anisotropy leads to a shift of the minimum in the c-fermion spectrum away from the $K$-point as is apparent in the MFT spectrum Fig. \ref{fig:minex_aniso}.
The TM spectra, Fig. \ref{fig:TM_aniso_p010p035a11}, do not display such a shift of the eigenvalues at the $K$-point (green $+$), neither for $\phi/\pi=0.1$ nor for $0.35$.
Furthermore, the symmetry $\Omega_{\rm min.}(\bk\pm{\bf K})=\Omega_{\rm min.}(\bk)$ is unaffected by the anisotropy, which is consistent with the MFT prediction, App. \ref{sec:appMFT}, for the $b$ and the $b+c$ fermion spectrum.

\section{Majorana MFT details}
\label{sec:appMFT}

Even before solving equations (\ref{eq:AB}) and (\ref{eq:BA}), it is
important to characterize the behavior of the Majorana fermions under
\emph{any} configuration of $A_{ij}$ and $B_{ij}$. The $c_i$ operators
describe fermions which move about the whole lattice with hopping
amplitudes given by $A_{ij}$. Similarly, the $b_i^\a$ operators
describe fermion hopping with amplitudes $B_{ij}K_{ij}^{\a\b}$. Most
importantly, the structure of $K_{ij}^{\a\b}$, given in
Eq. (\ref{eq:Kdef}), separates the $b_i^\a$ fermions into three
independent, uncorrelated sectors. Each sector is associated with one
of the three sublattices of hexagons. Within each sector, the hopping
amplitude is $B_{ij}\G$ around a hexagon of the corresponding
sublattice, and $B_{ij}K$ between neighboring hexagons. When $\G=0$,
$b_i^\a$ fermions are bound to a bond, while they are bound to a
hexagon when $K=0$. This behavior echoes the macroscopic degeneracies
of the parent classical models: in the classical Kitaev model there
are a macroscopic number of degenerate ground states which are related
to each other by reversing the sign of a single spin component for two
spins on the same bond. The classical $\G$-model has a similar
macroscopic degeneracy, obtained by reversing the signs of one
component of each of the six spins on one
hexagon\cite{rousochatzakis_classical_2017}.  Furthermore, returning
to the uniform mean-field solution, the $b$-fermion band structure is
similar to the band structure obtained in a Luttinger-Tisza study of
the corresponding classical models, since both are determined by the
exchange matrix $K_{ij}^{\a\b}$. In the classical case, the flat bands
for $\G = 0$ or $K = 0$ indicate the existence of a macroscopic number
of ground states, mentioned above.

\begin{figure}[t]
  \centering
  \includegraphics[width=\linewidth]{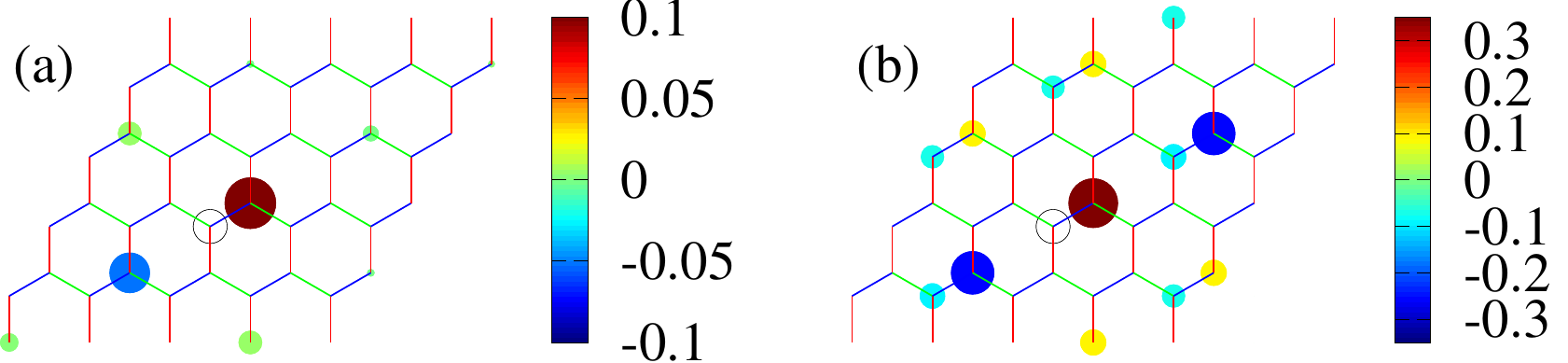}
  \caption{Spin-spin correlations in real space. The open black
    circles indicate site $i=0$ and the filled circles indicate
    $i$. The circle size and color represent the magnitude and value
    of the corresponding correlation
    $\braket{S_i^xS_0^x}$. (a) Mean-field results for
    $\phi/\pi = 0.4$. (b) ED results for $\phi/\pi = 0.5$. In both
    cases one finds same spin component correlations only among a
    subset of all sites.}
  \label{fig:SzSz}
\end{figure}

Spin correlations in real space, shown in Fig. \ref{fig:SzSz}a, reveal
a pattern which manifests the separation of $b$ fermions into
independent sectors: $S_i^z$ on site $i$ is correlated only with
a certain subset of $S_j^z$'s on other sites $j$. Similar
patterns are also obtained with ED, see Fig \ref{fig:SzSz}b although
some differences should be pointed out. According to the MFT, at
$\phi=0.5$ ($K=0$), there are non-zero spin-spin correlation only
within the same hexagons since the $b$-fermions are localized. Thus,
both finite $K$ and $\G$ are requried to get longer range
correlations. Furthermore, the correlation in Fig. \ref{fig:SzSz}b
don't decay very fast since they were obtained using ED on a small
cluster with periodic boundary condition, whereas the correlations in
\ref{fig:SzSz}a where calculating assuming an infinite
system. Finally, even among the the subset of sites which have same
spin component correlations, MFT gives nonzero \emph{static}
correlations only between opposite sublattice sites.

\begin{figure}[t]
  \centering
  \includegraphics[width=\linewidth]{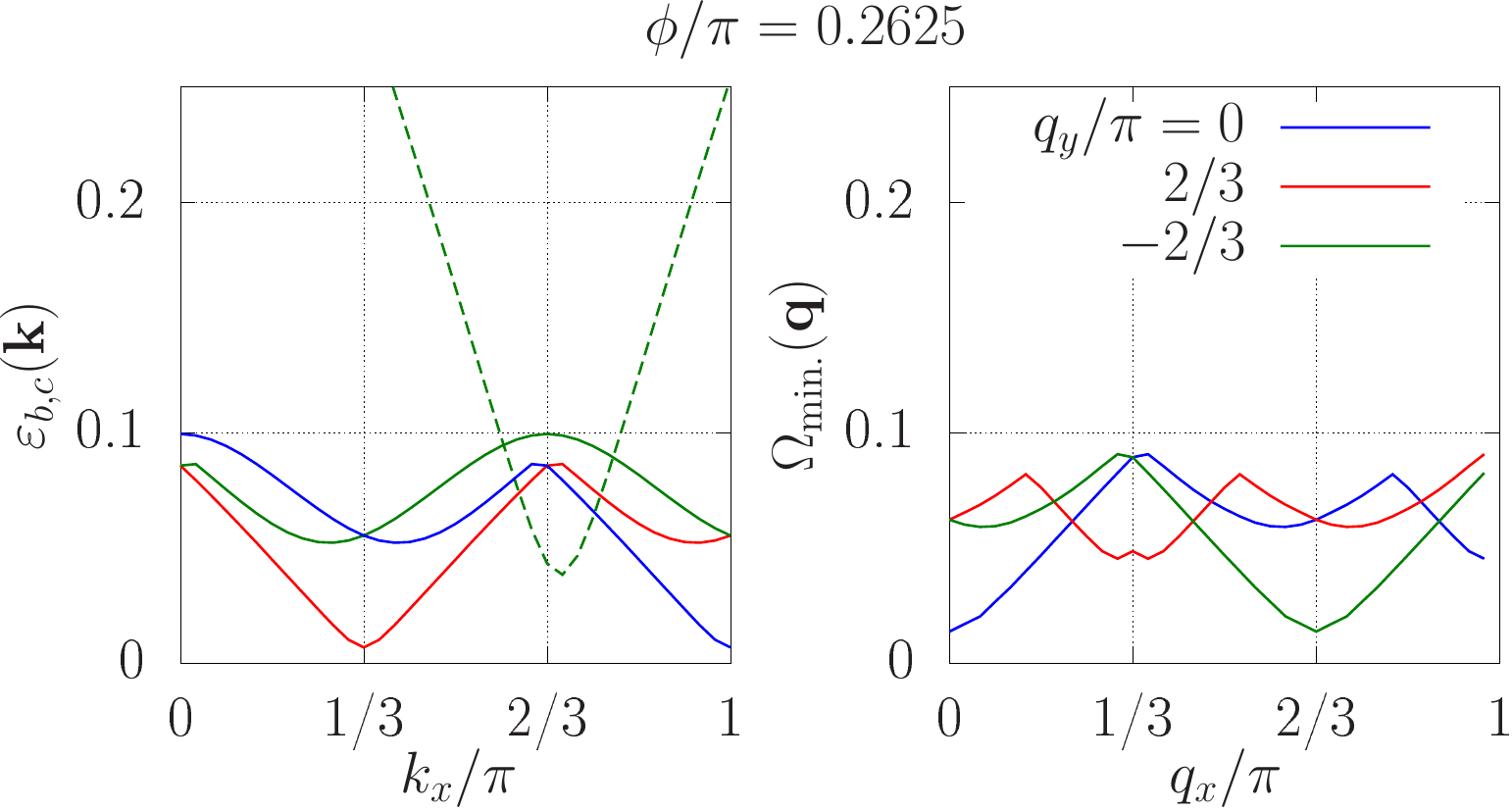}
  \caption{Left: $\varepsilon_c(\bk)$ (dashed) and
    $\varepsilon_b(\bk)$ (solid) for anisotropic mean-field amplitudes
    $A_x=1,A_y=1.1,A_z=1.2$ and $B_x=0.45,B_y=0.5,B_z=0.55$. Right:
    minimum energy for $b+c$ excitations. The spectra are plotted
    along the allowed momentum cuts, as in Fig. \ref{fig:bz}.}
  \label{fig:minex_aniso}
\end{figure}
\begin{figure}[t]
  \centering
  \includegraphics[width=\linewidth]{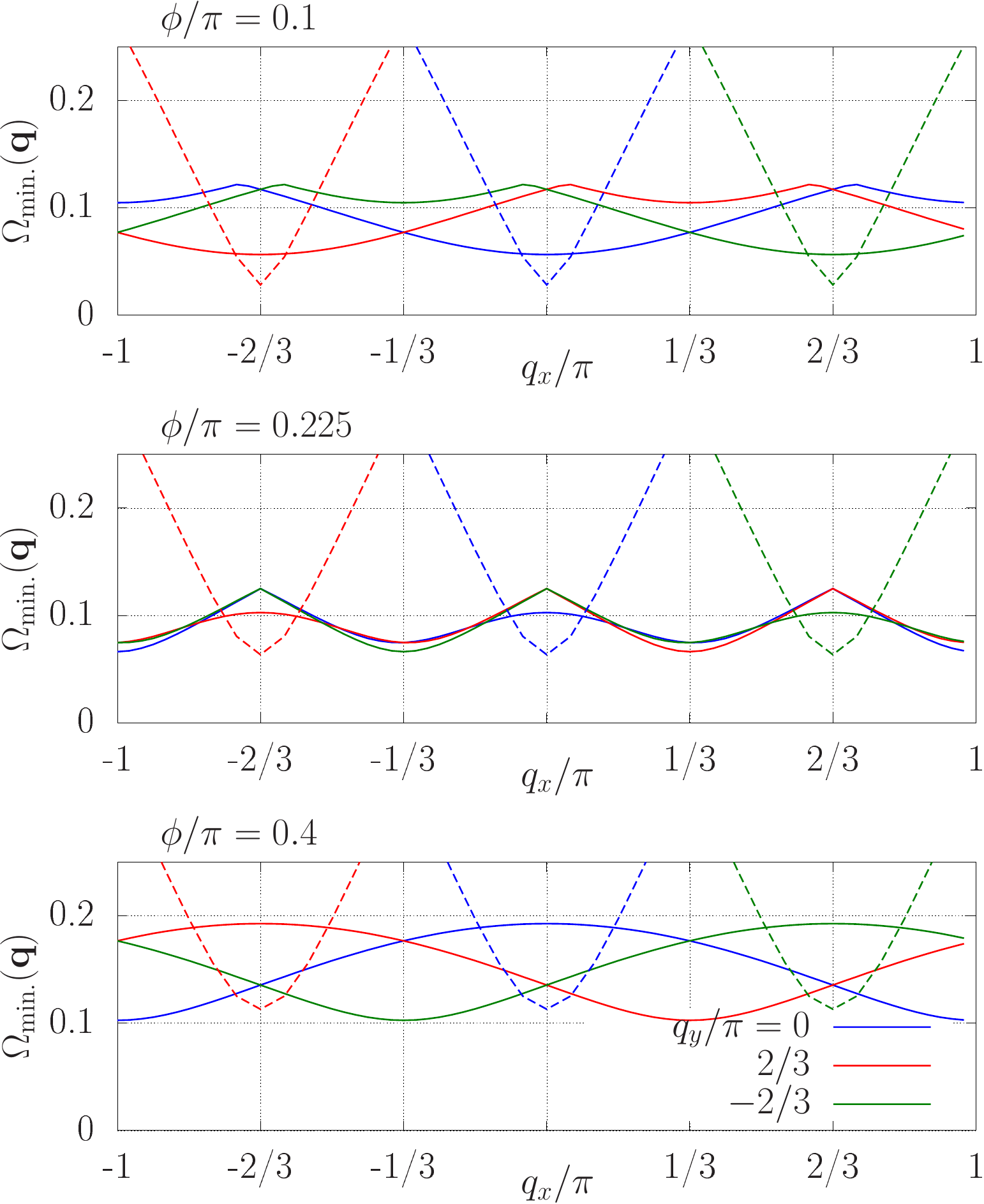}
  \caption{$\Omega_{\rm min.}(\bq)$ for $c+c$ (dashed) and $b+c$
    (solid) excitations, plotted along the momentum cuts in
    Fig. \ref{fig:bz} for different values of $\phi$.}
  \label{fig:minex_phi}
\end{figure}
Next we show how the symmetry $\varepsilon_b(\bk\pm{\bf K})$ emerges
from the Majorana mean-field Hamiltonian, Eq. (\ref{eq:Hmf}). We begin
by assuming that the mean-field amplitudes $B_{ij}$ preserve
translational invariance, and inversion symmetry, but are not
necessarily isotropic. We can therefore write the $b$-fermion Hamiltonian as 
\begin{equation}
  H_b=\sum_\bk\sum_{L,L'=A,B}\sum_{\alpha\beta}K_{LL'}^{\alpha\beta}(\bk)
  ib_L^\alpha(\bk)b_{L'}^\beta(-\bk),
\end{equation}
where the Fourier transform of the $b$ Majorana operators is
\begin{equation}
  \label{eq:bfourier}
  b_i=\frac{1}{\sqrt{N}}\sum_{\bk}b_L(\bk)e^{i\bk\cdot{\bf r}_i},
\end{equation}
and where $L=A,B$ denotes the sublattice of site $i$. The matrix
$K_{LL'}^{\alpha\beta}(\bk)$ is the Fourier transform of
$B_{ij}K_{ij}^{\alpha\beta}$, with $K_{ij}^{\alpha\beta}$ defined in
Eq. (\ref{eq:Kdef}). Thus, the sublattice off-diagonal block take the
form
\begin{equation}
  \label{eq:KABk}
  K_{AB}(\bk) = \left(
    \begin{array}{ccc}
      KB_xT_1 & \G B_z & \G B_yT_2 \\
      \G B_z & KB_yT_2 & \G B_xT_1 \\
      \G B_yT_2 & \G B_xT_1 & K B_z
    \end{array}\right),
\end{equation}
where $B_{x,y,z}$ are the anisotropic mean-field amplitudes, and
$T_i=e^{i\bk\cdot{\bf a}_i}$, with ${\bf a}_{1,2}$ the primitive
lattice vectors for the honeycomb lattice. $K$ itself is given by
\begin{equation}
  \label{eq:Kk}
  K(\bk) = \left(
    \begin{array}{cc}
       & K_{AB}(\bk) \\ K_{AB}^\dagger(\bk) & 
    \end{array}\right).
\end{equation}
Under a shift in momentum $\bk\to\bk\pm{\bf K}$, $T_1\to T_1\eta$ and
$T_2\to T_3\eta^*$, where $\eta=e^{2\pi i/3}$.  Using the phases
$\eta$ we next observe that
\begin{equation}
  \label{eq:KABkshift}
  K_{AB}(\bk) = \left(
    \begin{array}{ccc}
      1 & & \\ & \eta^* & \\ & & \eta
    \end{array}\right)
  K_{AB}(\bk\pm{\bf K})\left(
    \begin{array}{ccc}
      \eta & & \\ & 1 & \\ & & \eta^*
    \end{array}\right).
\end{equation}
The above relation shows that there exists a unitary transformation
which takes $K(\bk\pm{\bf K})\to K(\bk)$, and therefore, the spectrum
at the shifted momentum must be the same. As noted in the main text,
this symmetry holds irrespective of the details of the $c$ fermion
spectrum, and is therefore always induced in the $b+c$ spectrum as
well. Fig. \ref{fig:minex_aniso} shows the single and two particle
spectrum with fully anisotropic mean-field couplings. The minimum in
the $c$ fermions spectrum is clearly shifted away from the $K$ point
$(2\pi/3,-2\pi/3)$, while the $b+c$ spectrum still exhibits the
symmetry $\Omega_{\rm min.}(\bk\pm{\bf K})=\Omega_{\rm min.}(\bk)$.
Finally, in Fig. \ref{fig:minex_phi} we use the mean-field
Hamiltonian, Eq. (\ref{eq:Hmf}), to demonstrate, that the soft modes
in the two-particle spectrum move in momentum space, as $\phi$ is
increased.

\section{Dynamical structure factors and the Krylov subspace method}
\label{sec:appDSF}
The dynamical structure factor may be written as 

\begin{equation}
S(\vec{Q},\omega)
=
\sum_{\alpha=x,y,z}
S^{\alpha\alpha}(\vec{Q},\omega),
\end{equation}
where
\begin{equation}
S^{\alpha\alpha}(\vec{Q},\omega) = 
-\frac{1}{\pi}{\rm Im}
\bra{0}\hat{S}_{-\vec{Q}}^{\alpha}\frac{1}{\omega+i\delta-\hat{H}+E_0}\hat{S}_{+\vec{Q}}^{\alpha}\ket{0},
\label{originalSQomega}
\end{equation}
and
$\vec{\hat{S}}_{+\vec{Q}}=(\hat{S}_{+\vec{Q}}^{x},\hat{S}_{+\vec{Q}}^{y},\hat{S}_{+\vec{Q}}^{z})$
is the Fourier transform of the spin operators $\vec{\hat{S}}_i$
defined as,
\begin{equation} \vec{\hat{S}}_{\vec{Q}}=N^{-1/2}\sum_{i}^{N}\vec{\hat{S}}_i
  e^{+i\vec{Q}\cdot\vec{r}_i}.
\end{equation} The dynamical structure factors are conventionally
calculated by using the Lanczos algorithm initialized with an excited
state $\hat{S}_{+\vec{Q}}^{\alpha}\ket{0}$~\cite{gagliano1987,RevModPhys.66.763}.
However, naive implementations of the Lanczos algorithm and continued fraction\cite{gagliano1987} requires careful examination of the convergence of excited states relevant to the spectra to control the truncation errors.
In contrast, modern Krylov subspace
methods, which extract essence from the Lanczos algorithm, offer
controlled convergence without additional numerical costs.

Here, we solve a linear equation by employing a conjugate gradient
(CG) method, instead of explicitly calculating the resolvent of
$\hat{H}$ in Eq.(\ref{originalSQomega}).  The CG methods find the
solution in a Krylov subspace, as follows.  First, by introducing the
following two vectors,
\begin{eqnarray} \ket{\chi (\zeta)}
&=&(\zeta-\hat{H})^{-1}\hat{S}_{+\vec{Q}}^{\alpha}\ket{0},
\\ \ket{\phi} &=& \hat{S}_{+\vec{Q}}^{\alpha}\ket{0}
\end{eqnarray}

we rewrite $S^{\alpha\alpha}(\vec{Q},\omega)$ as
\begin{equation} S^{\alpha\alpha}(\vec{Q},\omega) =-\frac{1}{\pi}{\rm
Im} \braket{\phi|\chi(\omega+i\delta)}.
\end{equation} To obtain the unknown vector $\ket{\chi(\zeta)}$, we
solve the following linear equation,
\begin{equation} (\zeta-\hat{H})\ket{\chi (\zeta)} =
\ket{\phi}.\label{linearSQomega}
\end{equation} When the linear dimension of the matrix $\hat{H}$,
$\mathcal{L}$, is too large to store the whole matrix in the memory,
the linear equation is solved iteratively, for example, by using the
CG methods.  At $n$th iteration, the conjugate gradient algorithm
initialized with $\ket{\chi_0(\zeta)}=\ket{\phi}$ finds an approximate
solution $\ket{\chi_n (\zeta)}$ within a $n$-dimensional Krylov
subspace $\mathcal{K}_n (\zeta-\hat{H},\ket{\phi})={\rm
span}\{\ket{\phi},(\zeta-\hat{H})\ket{\phi},\dots,
(\zeta-\hat{H})^{n-1}\ket{\phi}\}$.  At each steps, the CG-type
algorithms search the approximate solution $\ket{\chi_n (\zeta)}$ to
minimize the 2-norm of the residual vector,
\begin{equation} \ket{\rho_n (\zeta)}= (\zeta-\hat{H})\ket{\chi_n
(\zeta)} - \ket{\phi}.
\end{equation} We note that one needs to solve
Eq.(\ref{linearSQomega}) essentially once at a fixed complex number
$\zeta=\omega+i\delta$ to obtain whole spectrum
$S^{\alpha\alpha}(\vec{Q},\omega)$.  Due to the shift invariance of
the Krylov subspace~\cite{frommer2003bicgstab}, namely, $\mathcal{K}_n
(\zeta-\hat{H},\ket{\phi})=\mathcal{K}_n (\zeta'-\hat{H},\ket{\phi})$
for any complex number $\zeta'\neq \zeta$, we can obtain $\ket{\chi
(\zeta')}$ from $\ket{\chi (\zeta)}$ with a numerical complexity of
$\mathcal{O}(\mathcal{L}^0)$~\cite{frommer2003bicgstab}.  The Krylov
subspace methods utilizing the shift invariance are called the shifted
Krylov subspace methods.

For the calculations of $S(\vec{Q},\omega)$, we employ the shifted
biconjugate gradient (BiCG) method implemented in a numerical library
$K\omega$ for the shifted Krylov subspace method~\cite{Komega}.  The
condition for truncating the shifted BiCG iteration is set
$\underset{\omega}{{\rm max}}\{\|\ket{\rho_n
  (\omega+i\delta)}\|\}<10^{-4}$ for the following calculations with
$\delta =0.02$.  The number of the iteration steps for satisfying the
condition depends on the parameters ($\phi$, $J_3$, and $B$), and is
typically of the order of one thousand and at most of the order of ten
thousand.

\bibliography{kgamma}

\end{document}